\documentclass[twocolumn, trackchanges]{aastex7}

\usepackage{CJK}
\usepackage{makecell}
\usepackage[multiple]{footmisc}
\newcommand{\corr}[1]{{\color{black}#1}}
\newcommand{\N}[1]{\textcolor{blue}{ #1}}

\newcommand{\CfA}{Center for Astrophysics $|$ Harvard \& Smithsonian, Cambridge, MA 02138, USA}
\newcommand{\IAIFI}{The NSF AI Institute for Artificial Intelligence and Fundamental Interactions}
\newcommand{\MIT}{Department of Physics and Kavli Institute for Astrophysics and Space Research, Massachusetts Institute of Technology, 77 Massachusetts Avenue, Cambridge, MA 02139, USA}

\usepackage{xcolor}

\begin{document}

\title{Enabling Early Transient Discovery in LSST via Difference Imaging with DECam}


\begin{CJK*}{UTF8}{gbsn}
\author[0000-0002-7937-6371,sname='Yize',gname='Dong']{Yize Dong (董一泽)}
\affiliation{\CfA}
\affiliation{\IAIFI}
\email[show]{yize.dong@cfa.harvard.edu} 

\author[orcid=0000-0002-9886-2834,sname='de~Soto']{Kaylee de~Soto}
\affiliation{\CfA}
\email{kaylee.de_soto@cfa.harvard.edu} 

\author[0000-0002-5814-4061,sname='Victoria Ashley',gname='Villar']{V.~Ashley Villar}
\affiliation{\CfA}
\affiliation{\IAIFI}
\email{ashleyvillar@cfa.harvard.edu}

\author[0000-0002-2028-9329, sname='Nugent']{Anya Nugent}
\affiliation{\CfA}
\email{anya.nugent@cfa.harvard.edu}

\author[0000-0003-4906-8447, sname='Gagliano']{Alex Gagliano}
\affiliation{\IAIFI}
\affiliation{\CfA}
\affiliation{\MIT}
\email{gaglian2@mit.edu}

\author[0000-0002-4924-444X, sname='Bostroem', gname='K.\ Azalee']{K.\ Azalee Bostroem}
\altaffiliation{LSSTC Catalyst Fellow}
\affiliation{Steward Observatory, University of Arizona, 933 North Cherry Avenue, Tucson, AZ 85721-0065, USA}
\email{abostroem@gmail.com}


\author[0009-0000-7835-3963,gname='Anastasia',sname='Alexov']{Anastasia~Alexov}
\affiliation{Vera C.\ Rubin Observatory Project Office, 950 N.\ Cherry Ave., Tucson, AZ  85719, USA}
\email{aalexov@lsst.org}

\author[0000-0002-5592-023X,gname='Éric',sname='Aubourg']{\'{E}ric~Aubourg}
\affiliation{Universit\'{e} Paris Cit\'{e}, CNRS/IN2P3, CEA, APC, 4 rue Elsa Morante, F-75013 Paris, France}
\email{eric@aubourg.net}


\author[0000-0002-4052-2511,gname='Farrukh',sname='Azfar']{Farrukh~Azfar}
\affiliation{Department of Physics, University of Oxford, Denys Wilkinson Building, Keble Road, Oxford, OX1 3RH, UK}
\email{farrukh.azfar@physics.ox.ac.uk}

\author[0000-0001-7387-2633,gname='Alexandre',sname='Boucaud']{Alexandre~Boucaud}
\affiliation{Universit\'{e} Paris Cit\'{e}, CNRS/IN2P3, APC, 4 rue Elsa Morante, F-75013 Paris, France}
\email{aboucaud@apc.in2p3.fr}

\author[gname='Andrew',sname='Bradshaw']{Andrew~Bradshaw}
\affiliation{SLAC National Accelerator Laboratory, 2575 Sand Hill Rd., Menlo Park, CA 94025, USA}
\affiliation{Kavli Institute for Particle Astrophysics and Cosmology, SLAC National Accelerator Laboratory, 2575 Sand Hill Rd., Menlo Park, CA 94025, USA}
\email{andrewkbradshaw@gmail.com}

\author[0000-0001-9022-4232,gname='Johann',sname='Cohen-Tanugi']{Johann~Cohen-Tanugi}
\affiliation{Universit\'{e} Clermont Auvergne, CNRS/IN2P3, LPCA, 4 Avenue Blaise Pascal, F-63000 Clermont-Ferrand, France}
\email{johann.cohentanugi@gmail.com}

\author[0000-0003-1131-7030,gname='Sylvie',sname='Dagoret-Campagne']{Sylvie~Dagoret-Campagne}
\affiliation{Universit\'{e} Paris-Saclay, CNRS/IN2P3, IJCLab, 15 Rue Georges Clemenceau, F-91405 Orsay, France}
\email{sylvie.dagoret-campagne@ijclab.in2p3.fr}

\author[gname='Philip',sname='Daly']{Philip~Daly}
\affiliation{Steward Observatory, University of Arizona, 933 North Cherry Avenue, Tucson, AZ 85721-0065, USA}
\email{pndaly@arizona.edu}

\author[gname='Felipe',sname='Daruich']{Felipe~Daruich}
\affiliation{Vera C.\ Rubin Observatory, Avenida Juan Cisternas \#1500, La Serena, Chile}
\email{fdaruich@lsst.org}

\author[gname='Peter E.',sname='Doherty']{Peter~E.~Doherty}
\affiliation{Smithsonian Astrophysical Observatory, 60 Garden St., Cambridge MA 02138, USA}
\email{peter.doherty@cfa.harvard.edu}

\author[0000-0002-7790-9971,gname='Holger',sname='Drass']{Holger~Drass}
\affiliation{Vera C.\ Rubin Observatory, Avenida Juan Cisternas \#1500, La Serena, Chile}
\email{hdrass@lsst.org}


\author[0000-0003-2933-391X,gname='Orion',sname='Eiger']{Orion~Eiger}
\affiliation{SLAC National Accelerator Laboratory, 2575 Sand Hill Rd., Menlo Park, CA 94025, USA}
\affiliation{Kavli Institute for Particle Astrophysics and Cosmology, SLAC National Accelerator Laboratory, 2575 Sand Hill Rd., Menlo Park, CA 94025, USA}
\email{eiger@slac.stanford.edu}

\author[0000-0003-0800-8755,gname='Leanne P.',sname='Guy']{Leanne~P.~Guy}
\affiliation{Vera C.\ Rubin Observatory, Avenida Juan Cisternas \#1500, La Serena, Chile}
\email{leanne.guy@noirlab.edu}

\author[gname='Patrick A.',sname='Hascall']{Patrick~A.~Hascall}
\affiliation{SLAC National Accelerator Laboratory, 2575 Sand Hill Rd., Menlo Park, CA 94025, USA}
\email{hascallp@gmail.com}

\author[0000-0001-5250-2633,gname='Željko',sname='Ivezić']{\v{Z}eljko~Ivezi\'{c}}
\affiliation{Vera C.\ Rubin Observatory Project Office, 950 N.\ Cherry Ave., Tucson, AZ  85719, USA}
\affiliation{University of Washington, Dept.\ of Astronomy, Box 351580, Seattle, WA 98195, USA}
\email{zivezic@lsst.org}

\author[gname='Fabrice',sname='Jammes']{Fabrice~Jammes}
\affiliation{Universit\'{e} Clermont Auvergne, CNRS/IN2P3, LPCA, 4 Avenue Blaise Pascal, F-63000 Clermont-Ferrand, France}
\email{fabrice.jammes@gmail.com}

\author[0000-0002-5751-3697,gname='M. James',sname='Jee']{M.~James~Jee}
\affiliation{Department of Astronomy, Yonsei University, 50 Yonsei-ro, Seoul 03722, Republic of Korea}
\affiliation{Physics Department, University of California, One Shields Avenue, Davis, CA 95616, USA}
\email{mkjee@yonsei.ac.kr}

\author[0000-0001-5982-167X,gname='Tim',sname='Jenness']{Tim~Jenness}
\affiliation{Vera C.\ Rubin Observatory Project Office, 950 N.\ Cherry Ave., Tucson, AZ  85719, USA}
\email{tjenness@lsst.org}

\author[0000-0003-4833-9137,gname='Steven M.',sname='Kahn']{Steven~M.~Kahn}
\affiliation{Physics Department,  University of California, 366 Physics North, MC 7300 Berkeley, CA 94720, USA}
\email{stevkahn@berkeley.edu}

\author[0000-0002-5261-5803,gname='Yijung',sname='Kang']{Yijung~Kang}
\affiliation{Kavli Institute for Particle Astrophysics and Cosmology, SLAC National Accelerator Laboratory, 2575 Sand Hill Rd., Menlo Park, CA 94025, USA}
\affiliation{Vera C.\ Rubin Observatory, Avenida Juan Cisternas \#1500, La Serena, Chile}
\email{ykang@slac.stanford.edu}

\author[0000-0001-9395-4759,gname='Lee S.',sname='Kelvin']{Lee~S.~Kelvin}
\affiliation{Department of Astrophysical Sciences, Princeton University, Princeton, NJ 08544, USA}
\email{lkelvin@astro.princeton.edu}

\author[gname='Ivan V.',sname='Kotov']{Ivan~V.~Kotov}
\affiliation{Brookhaven National Laboratory, Upton, NY 11973, USA}
\email{kotov@bnl.gov}


\author[0000-0003-1779-775X,gname='Gábor',sname='Kovács']{G\'abor~Kov\'acs}
\affiliation{Institute for Data-intensive Research in Astrophysics and Cosmology, University of Washington, 3910 15th Avenue NE, Seattle, WA 98195, USA}
\email{kgabor79@gmail.com}

\author[0000-0001-7178-8868,gname='Laurent',sname='Le Guillou']{Laurent~Le~Guillou}
\affiliation{Sorbonne Universit\'{e}, Universit\'{e} Paris Cit\'{e}, CNRS/IN2P3, LPNHE, 4 place Jussieu, F-75005 Paris, France}
\email{llg@lpnhe.in2p3.fr}

\author[gname='Shuang',sname='Liang']{Shuang~Liang}
\affiliation{SLAC National Accelerator Laboratory, 2575 Sand Hill Rd., Menlo Park, CA 94025, USA}
\email{sliang92@stanford.edu}

\author[0000-0003-2866-3802,gname='Mostafa',sname='Lutfi']{Mostafa~Lutfi}
\affiliation{Vera C.\ Rubin Observatory Project Office, 950 N.\ Cherry Ave., Tucson, AZ  85719, USA}
\email{mlutfi@lsst.org}

\author[gname='Morgan',sname='May']{Morgan~May}
\affiliation{Department of Physics Columbia University, New York, NY 10027, USA}
\affiliation{Brookhaven National Laboratory, Upton, NY 11973, USA}
\email{mm21@columbia.edu}

\author[0000-0001-6013-1131,gname='Guillem',sname='Megias Homar']{Guillem~Megias~Homar}
\affiliation{Division of Physics, Mathematics and Astronomy, California Institute of Technology, Pasadena, CA 91125, USA}
\email{gmegias@caltech.edu}

\author[gname='Marc',sname='Moniez']{Marc~Moniez}
\affiliation{Universit\'{e} Paris-Saclay, CNRS/IN2P3, IJCLab, 15 Rue Georges Clemenceau, F-91405 Orsay, France}
\email{marc.moniez@ijclab.in2p3.fr}


\author[gname='Freddy',sname='Muñoz Arancibia']{Freddy~Mu\~noz~Arancibia}
\affiliation{Vera C.\ Rubin Observatory Project Office, 950 N.\ Cherry Ave., Tucson, AZ  85719, USA}
\email{fmunoz@lsst.org}

\author[0000-0003-3827-4691,gname='Erfan',sname='Nourbakhsh']{Erfan~Nourbakhsh}
\affiliation{Department of Astrophysical Sciences, Princeton University, Princeton, NJ 08544, USA}
\email{erfan@astro.princeton.edu}

\author[0000-0002-7295-2743,gname='Hye Yun',sname='Park']{Hye~Yun~Park}
\affiliation{Department of Physics, Duke University, Durham, NC 27708, USA}
\email{hp175@duke.edu}

\author[0000-0001-5471-9609,gname='John R.',sname='Peterson']{John~R.~Peterson}
\affiliation{Department of Physics and Astronomy, Purdue University, 525 Northwestern Ave., West Lafayette, IN  47907, USA}
\email{peters11@purdue.edu}

\author[0000-0002-2598-0514,gname='Andrés A.',sname='Plazas Malagón']{Andr\'es~A.~Plazas~Malag\'on}
\affiliation{SLAC National Accelerator Laboratory, 2575 Sand Hill Rd., Menlo Park, CA 94025, USA}
\affiliation{Kavli Institute for Particle Astrophysics and Cosmology, SLAC National Accelerator Laboratory, 2575 Sand Hill Rd., Menlo Park, CA 94025, USA}
\email{plazas@stanford.edu}

\author[0000-0001-7445-4724,gname='Daniel',sname='Polin']{Daniel~Polin}
\affiliation{Physics Department, University of California, One Shields Avenue, Davis, CA 95616, USA}
\email{dapolin@ucdavis.edu}

\author[0000-0002-1557-3560,gname='Bruno C.',sname='Quint']{Bruno~C.~Quint}
\affiliation{Vera C.\ Rubin Observatory Project Office, 950 N.\ Cherry Ave., Tucson, AZ  85719, USA}
\email{bquint@lsst.org}

\author[0000-0002-0138-1365,gname='Tiago',sname='Ribeiro']{Tiago~Ribeiro}
\affiliation{Vera C.\ Rubin Observatory Project Office, 950 N.\ Cherry Ave., Tucson, AZ  85719, USA}
\email{tribeiro@lsst.org}

\author[0000-0001-8239-3079,gname='Vincent J.',sname='Riot']{Vincent~J.~Riot}
\affiliation{Lawrence Livermore National Laboratory, 7000 East Avenue, Livermore, CA 94550, USA}
\email{riot1@llnl.gov}

\author[0000-0002-9641-4552,gname='Cécile',sname='Roucelle']{C\'{e}cile~Roucelle}
\affiliation{Universit\'{e} Paris Cit\'{e}, CNRS/IN2P3, APC, 4 rue Elsa Morante, F-75013 Paris, France}
\email{roucelle@apc.in2p3.fr}

\author[0000-0002-8687-0669,gname='Bruno O.',sname='Sánchez']{Bruno~O.~S\'anchez}
\affiliation{Aix Marseille Universit\'{e}, CNRS/IN2P3, CPPM, 163 avenue de Luminy, F-13288 Marseille, France}
\email{bsanchez@cppm.in2p3.fr}

\author[0000-0002-9238-9521,gname='David',sname='Sanmartim']{David~Sanmartim}
\affiliation{Vera C.\ Rubin Observatory, Avenida Juan Cisternas \#1500, La Serena, Chile}
\email{dsanmartim@lsst.org}

\author[gname='Jacques',sname='Sebag']{Jacques~Sebag}
\affiliation{Vera C.\ Rubin Observatory, Avenida Juan Cisternas \#1500, La Serena, Chile}
\email{jsebag@lsst.org}

\author[0000-0003-4734-2019,gname='Nima',sname='Sedaghat']{Nima~Sedaghat}
\affiliation{University of Washington, Dept.\ of Astronomy, Box 351580, Seattle, WA 98195, USA}
\email{nimaseda@uw.edu}

\author[0000-0003-4058-5202,gname='Richard A.',sname='Shaw']{Richard~A.~Shaw}
\affiliation{Space Telescope Science Institute, 3700 San Martin Drive, Baltimore, MD 21218, USA}
\email{shaw@stsci.edu}

\author[0009-0000-6778-7168,gname='Alysha',sname='Shugart']{Alysha~Shugart}
\affiliation{Vera C.\ Rubin Observatory, Avenida Juan Cisternas \#1500, La Serena, Chile}
\email{alysha.shugart@noirlab.edu}

\author[0009-0001-6379-3365,gname='Ioana',sname='Sotuela Elorriaga']{Ioana~Sotuela~Elorriaga}
\affiliation{Vera C.\ Rubin Observatory, Avenida Juan Cisternas \#1500, La Serena, Chile}
\email{isotuela@lsst.org}

\author[0000-0002-9589-1306,gname='Krzysztof',sname='Suberlak']{Krzysztof~Suberlak}
\affiliation{University of Washington, Dept.\ of Astronomy, Box 351580, Seattle, WA 98195, USA}
\email{suberlak@uw.edu}

\author[0000-0001-9445-1846,gname='John D.',sname='Swinbank']{John~D.~Swinbank}
\affiliation{ASTRON, Oude Hoogeveensedijk 4, 7991 PD, Dwingeloo, The Netherlands}
\affiliation{Department of Astrophysical Sciences, Princeton University, Princeton, NJ 08544, USA}
\email{swinbank@astron.nl}

\author[0000-0002-9121-3436,gname='Sandrine',sname='Thomas']{Sandrine~Thomas}
\affiliation{Vera C.\ Rubin Observatory Project Office, 950 N.\ Cherry Ave., Tucson, AZ  85719, USA}
\email{sthomas@lsst.org}

\author[0000-0002-9242-8797,gname='J. Anthony',sname='Tyson']{J.~Anthony~Tyson}
\affiliation{Physics Department, University of California, One Shields Avenue, Davis, CA 95616, USA}
\email{tyson@lsst.org}

\author[0000-0002-1431-9245,gname='Wouter',sname='van Reeven']{Wouter~van~Reeven}
\affiliation{Vera C.\ Rubin Observatory, Avenida Juan Cisternas \#1500, La Serena, Chile}
\email{wvanreeven@lsst.org}

\author[0000-0002-4557-6682,gname='Charlotte',sname='Ward']{Charlotte~Ward}
\affiliation{Department of Astronomy and Astrophysics, The Pennsylvania State University, 525 Davey Lab, University Park, PA 16802, USA}
\email{cvw5890@psu.edu}

\author[0000-0003-1989-4879,gname='Christopher Z.',sname='Waters']{Christopher~Z.~Waters}
\affiliation{Department of Astrophysical Sciences, Princeton University, Princeton, NJ 08544, USA}
\email{czw@astro.princeton.edu}

\author[gname='Oliver',sname='Wiecha']{Oliver~Wiecha}
\affiliation{Vera C.\ Rubin Observatory Project Office, 950 N.\ Cherry Ave., Tucson, AZ  85719, USA}
\email{oliver@straightengineering.com}

\author[0000-0001-7113-1233,gname='W. M.',sname='Wood-Vasey']{W.~M.~Wood-Vasey}
\affiliation{Department of Physics and Astronomy, University of Pittsburgh, 3941 O'Hara Street, Pittsburgh, PA 15260, USA}
\email{wmwv@pitt.edu}



\begin{abstract}

We present \texttt{SLIDE}, a pipeline that enables transient discovery in data from the Vera C. Rubin Observatory's Legacy Survey of Space and Time (LSST), using archival images from the Dark Energy Camera (DECam) as templates for difference imaging. We apply this pipeline to the recently released Data Preview 1 (DP1; the first public release of Rubin commissioning data) and search for transients in the resulting difference images. The image subtraction, photometry extraction, and transient detection are all performed on the Rubin Science Platform. We demonstrate that \texttt{SLIDE} effectively extracts clean photometry by circumventing poor or missing LSST templates.
We identified 29 previously unreported transients, 12 of which would not have been detected based on the DP1 \texttt{DiaObject} catalog.
\texttt{SLIDE} will be especially useful for transient analysis in the early years of LSST, when template coverage will be largely incomplete or when templates may be contaminated by transients present at the time of acquisition. We present multiband light curves for a sample of known transients, along with new transient candidates identified through our search. Finally, we discuss the prospects of applying this pipeline during the main LSST survey. Our pipeline is broadly applicable and will support studies of all transients with slowly evolving phases.



\end{abstract}





\section{Introduction} 
The wide field of view and exceptional depth of the Vera C. Rubin Observatory will usher in a new era for time-domain astronomy \citep{LSST2009arXiv0912.0201L}. Expected to begin full operations in late 2025, the Rubin Observatory's decade-long Legacy Survey of Space and Time (LSST, \citealt{Ivezic2019ApJ...873..111I}) will stream up to ten million transient alerts nightly. 
Data taken between 4 November 2023 and 10 December 2024 with the LSST Commissioning Camera (``\mbox{LSSTComCam}''), which uses the same hardware as the LSST Camera but with a reduced field of view \citep{lsstcam}, were released as Data Preview 1 (DP1\footnote{\url{https://rtn-095.lsst.io}}; \citealt{RTN-095,DP1,10.71929/rubin/2561361}).

LSST's nightly alerts rely on difference imaging, which compares new observations to deep reference templates to identify brightness fluctuations from variable and transient sources.
Template collection is expected to continue through the first year of regular survey operations \citep{leanne_p_guy_2023_10059624,guy_RTN-011}. Transients observed during this early period may contaminate templates, making it difficult to accurately distinguish their flux from contaminating background light (e.g., from the host galaxy). 
\corr{Such contamination may not significantly affect the transient detections for rapidly-evolving transients or supernovae (SNe) that change brightness on relatively short timescales. }
Despite inaccurate reported photometry, these transients can still be detected by the LSST alert stream even if partially imprinted in the templates. 


In contrast, long-duration transients that do not exhibit strong luminosity evolution will be difficult to identify in real time if they are present in the LSST templates. \corr{This includes long-lived precursors to core-collapse SNe (CCSNe; \citealt{Pastorello2007Natur.447..829P,
Strotjohann2021ApJ...907...99S,
Tsuna2023,
Tsuna2024arXiv240102389T,
Brennan2025arXiv250308768B}) such as those detected in SN~2020tlf \citep{Jacobson-Galan2022ApJ...924...15J}, SN~2023fyq \citep{Dong2024ApJ...977..254D,Brennan2024A&A...684L..18B}, SN~2023zkd \citep{Gagliano2025arXiv250219469G}, as well as long-duration superluminous SNe \citep[e.g.,][]{Gomez2024MNRAS.535..471G} and luminous red novae (LRNe; \citealt{Mauerhan2015MNRAS.447.1922M,Smith2016MNRAS.458..950S,Blagorodnova2017ApJ...834..107B}). }
Even if such transients are detected, obtaining clean photometry will require waiting until they have faded, at which point templates can be constructed. This delay could hinder timely follow-up and downstream analysis during the early phases of the LSST survey.

The release of Rubin DP1 provides a valuable dataset for developing and testing infrastructure that can be applied to real LSST data. In this paper, we present the \texttt{SLIDE} package for performing LSST image subtraction using images taken by the Dark Energy Camera (DECam; \citealt{Honscheid2008arXiv0810.3600H,Flaugher2015AJ....150..150F}). This provides an alternative to LSST-derived templates, and will benefit the broader transient community in the initial years of the LSST survey.  

\corr{We have released our \texttt{SLIDE} package on GitHub\footnote{\url{https://github.com/yizedong/SLIDE}}.}
\texttt{SLIDE} is intended to be run directly on the Rubin Science Platform \citep{OMullane2024ASPC..535..227O,rsp2}. 
We use \texttt{SLIDE} to search for DP1 transients within two of the extragalactic fields observed by LSSTComCam: the Extended Chandra Deep Field South (ECDFS) and the Euclid Deep Field South (EDFS). 
\corr{We find that all transients reported to the Transient Name Server (TNS; \citealt{tns}) within our selected search area are successfully recovered, provided they had not already faded by the time the images were taken. }
In addition, we identify 29 previously unreported transients, 18 of which are likely nuclear transients, and 12 of which are either not present or have fewer than two detections in DP1's \texttt{DiaObject} catalog \citep{diaobject}.

In Section~\ref{sec:slide_sub}, we provide an overview of \texttt{SLIDE} and test it on a Rubin LSSTComCam transient with template contamination. In Section~\ref{sec:search}, we describe our search for transient candidates in the EDFS and ECDFS fields using our corrected difference images. We summarize our findings and outline future prospects with the full LSST data stream in Section~\ref{sec:conclusion}.

\section{LSST image subtraction with \texttt{SLIDE}}\label{sec:slide_sub}

\texttt{SLIDE} can be easily installed on the Rubin Science Platform\footnote{\url{https://ldm-542.lsst.io}}\textsuperscript{,}\footnote{\url{https://lse-319.lsst.io}} \citep{OMullane2024ASPC..535..227O}. We include an example notebook to demonstrate its usage \footnote{\url{https://github.com/yizedong/SLIDE/blob/main/example.ipynb}}. Here, we outline its major components.

DECam is a wide-field \corr{charge-coupled device (CCD)} imager mounted on the 4~m Blanco telescope at Cerro Tololo Inter-American Observatory (CTIO) in northern Chile. Initially designed for the Dark Energy Survey (DES, \citealt{DES2016MNRAS.460.1270D}), DECam consists of 62 science CCDs with a pixel scale of 0.263 arcsec/pixel and a field of view of approximately 3 $\rm deg^{2}$. The DES survey was conducted from 15 August 2013 to 9 January 2019 and covered 5000 $\rm deg^{2}$ in $grizY$ bands.

\texttt{SLIDE} automatically retrieves deep coadded images from the DES Data Release 2 (DR2; \citealt{Abbott2021ApJS..255...20A}) that overlap a position of interest. The final coadds reach a median 5$\sigma$ depth of $g = 25.4$, $r = 25.1$, $i = 24.5$, $z = 23.8$, and $Y=22.4$, which are deeper than single-visit depths expected from LSST and deeper than images released as part of DP1 \citep{Bianco2022ApJS..258....1B, DP1}. This makes DES DR2 images suitable as templates for LSST image subtraction. All the difference images used in this Letter are made using the DES DR2 templates.

Alternatively, \texttt{SLIDE} can retrieve coadded DECam images from the Dark Energy Camera Legacy Survey (DECaLS; \citealt{2016AAS...22831701B, DECaLS}) for use as templates. These images have similar or slightly greater depth than single exposures in DP1, making them a useful alternative when DES templates are not available. Users may also supply custom DECam templates.

DES DR2 (or DECaLS) images are retrieved using the Simple Image Access (SIA) service provided by the Astro Data Lab \citep{datalab1, datalab2}. DECam images are aligned and rescaled to match the LSST images using the \texttt{reproject} package\footnote{\url{https://github.com/astropy/reproject}} which uses an adaptive, anti-aliased resampling algorithm \citep{DeForest2004SoPh..219....3D}. The point spread function (PSF) of the DECam images is modeled using \texttt{Photutils} with stars selected from the Gaia DR3 catalog \citep{Gaia2016A&A...595A...1G,Gaia2021A&A...649A...1G,Gaia2021A&A...649A...3R}. 

For DP1 images, we obtain the calibrated exposure images \corr{(\texttt{visit\_image})} from the Rubin Science Platform using the Butler \citep{Jenness2022SPIE12189E..11J}. 
These images have been processed by the LSST Science Pipelines \citep{Rubin_Science_Pipeline} and are ready for scientific use. 
LSSTComCam consists of nine CCDs; \texttt{SLIDE} can operate on either full CCD images or cutout regions.
The LSSTComCam image PSFs provided can vary slightly over the field of view. Therefore, we use the median PSF of the detector for image subtraction. Alternatively, our package offers options to recalculate the PSF and refine the WCS of images using stars from Gaia DR3.

The image subtraction is performed using a Python implementation of the Zackay-Ofek-Gal-Yam (ZOGY, \citealt{Zackay2016ApJ...830...27Z, david_guevel_2017_1043973}) algorithm \footnote{\url{https://github.com/dguevel/PyZOGY}}, which provides mathematically optimal statistics for image subtraction and does not require that the reference image has sharper PSF than the science image. The runtime for PSF construction and image subtraction depends on image size: for a full LSST CCD image (4000$\times$4000 pixels), it takes approximately 1.5 minutes, while for a 1500$\times$1500 pixel cutout, it takes about 15 seconds.

Finally, PSF photometry is performed on the difference images at \corr{specified positions (RA and DEC)} using the \texttt{Photutils} package from \texttt{astropy} \citep{Astropy2022ApJ...935..167A}.

\begin{figure*}
    \includegraphics[width=1.\linewidth]{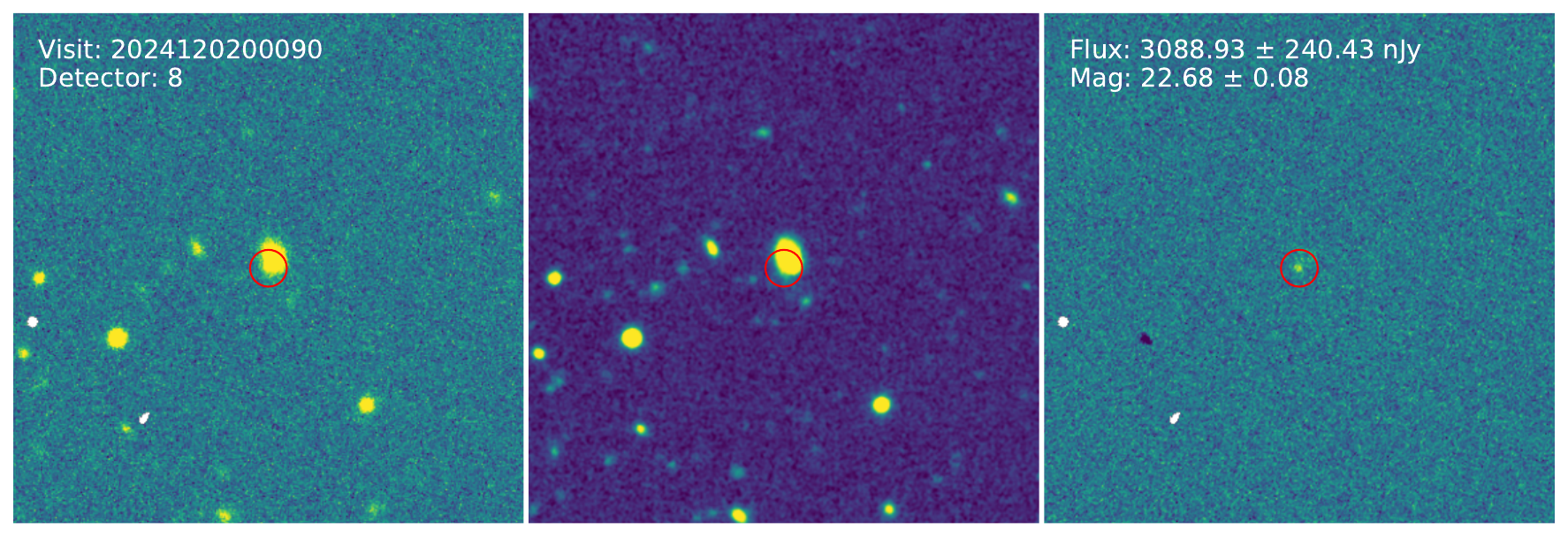}
    \includegraphics[width=1.\linewidth]{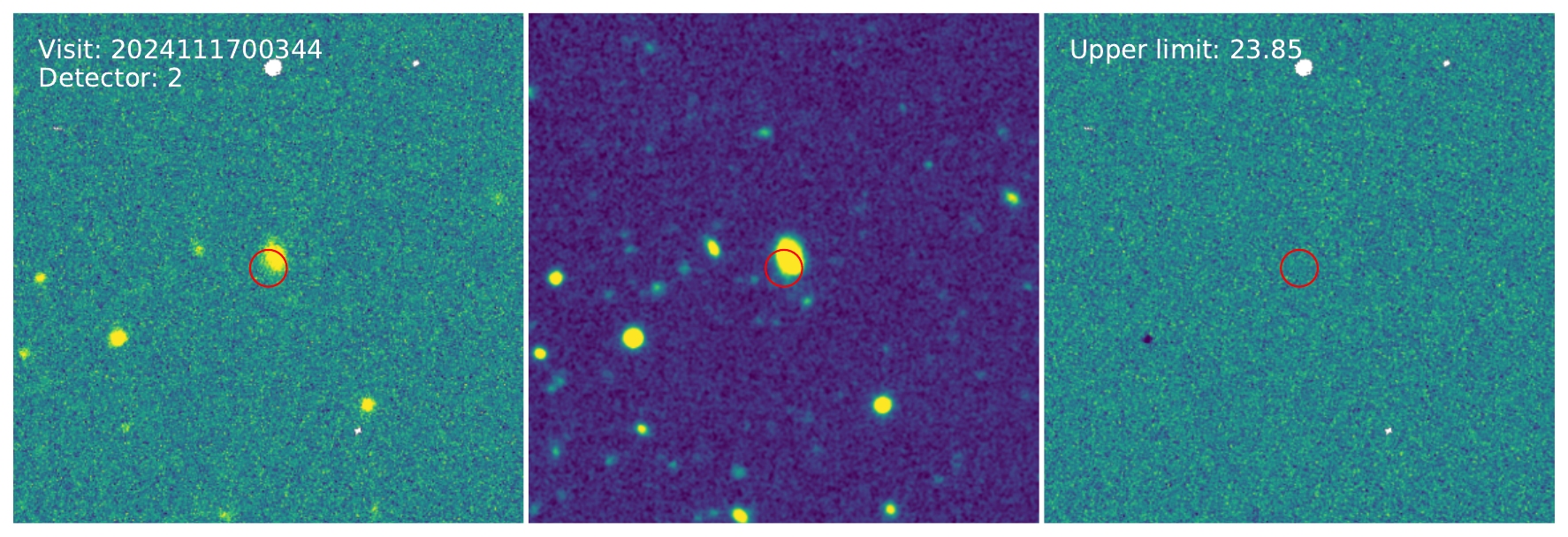}
    \caption{\textit{Upper}: Image subtraction using DECam templates for visit 2024120200090, detector 8 taken on 2 December 2024 \corr{in the $r$ band}. \corr{The image is oriented with North at the top and East to the left.} AT~2024ahzi is visible on the image and marked by the red circle. 
    The left and \corr{middle} panels show the LSST image and the coadded DECam image, respectively, \corr{while the right panel shows the difference image.}
    The displayed cutouts are 400$\times$400 pixels in size, centered on the transient position. The white patches are bad pixels and are masked out prior to subtraction.
    PSF photometry is performed at the location of AT~2024ahzi on the difference image (center), and the results are annotated on the panel.
    \textit{Lower}: Image subtraction for visit 2024111700344, detector \corr{2}, taken on 17 November 2024 \corr{in the $r$ band}. AT~2024ahzi is not visible on the difference image, and an upper limit is derived.
    \label{fig:2024ahzi}}
\end{figure*}

We test \texttt{SLIDE} on a known transient, AT~2024ahzi, which was reported to TNS on 13 March 2025 \citep{Murphey2025TNSTR.975....1M, Andreoni2025TNSAN.204....1A}. Flux from AT~2024ahzi is present in the LSSTComCam templates used by DP1, making the reported difference-imaged forced photometry unreliable \citep{de_soto_in_prep}. We process all available DES images overlapping the transient position and find that the resulting photometry is consistent with that obtained by the Young Supernova Experiment \citep[YSE;][]{2021Jones_YSE, 2023Aleo_YSEDR1} using DECam \corr{within $\sim1\sigma$} (see detailed photometric comparison in \citealt{de_soto_in_prep}). In Figure~\ref{fig:2024ahzi}, we show examples of image subtraction at the position of AT~2024ahzi using the DES templates as a demonstration of the subtraction quality. The subtractions are generally clean, and the transient is clearly detected in the center when present.

\section{Candidate Transients Search}\label{sec:search}

\subsection{Field Selection}\label{sec:fields}

Rubin DP1 contains $ugrizy$ images from seven fields, taken with LSSTComCam between November and December 2024. 
We select the ECDFS and the EDFS fields to perform an experimental transient search, as these are the most well-observed fields in DP1, are far from the Galactic plane, and have sufficiently overlapping coverage by DES. 
For each field, we select a subset of $r$-band exposures that maximize overall spatial overlap across visits while also ensuring even temporal coverage.
We restrict the transient search to a single filter to reduce the computational workload; we prefer the $r$-band as it has the highest cadence in both fields. This selection yields 37 visits of ECDFS and 18 visits of EDFS. For each visit, image subtraction is performed on each CCD's image independently.

\subsection{Transient Detection} \label{sec:selection}
Transient detection is performed on the difference images using \texttt{SEP} \citep{Barbary2016JOSS....1...58B}, a Python implementation of Source Extractor \citep{Bertin1996A&AS..117..393B}. 

Stars brighter than approximately 16 mag saturate LSSTComCam's 30-second exposures. Although these bright stars are masked out in the difference images, we find that they often produce prominent spike-like artifacts in the surrounding area, which can be misidentified as transient detections. To mitigate contamination, we exclude any candidates located within 20~arcseconds of such stars. 
\corr{Additionally, we find that stars with proper motion may be misaligned between the template and science images. Such misalignment can create residual artifacts in the difference images, and we therefore exclude candidates located within one full-width at half maximum (FWHM) of sources classified as stars in the DES DR2 catalog \citep{Abbott2021ApJS..255...20A}.}

To identify transients of interest from the remaining \corr{15,785} targets, we require that each candidate has at least 3 detections, \corr{which removes contamination from cosmic rays and other artifacts}. We also require that the peak-to-peak flux variation exceed three times the mean flux uncertainty, that the standard deviation of the flux exceed the mean flux uncertainty, and that the peak-to-peak magnitude variation be greater than 0.3 mag. Sources that do not meet all of these criteria are excluded from further analysis. 

We associate the remaining 1224 candidates to likely host galaxies using \texttt{Pr\"ost}\footnote{\url{https://github.com/alexandergagliano/Prost}} \citep{Gagliano2025_Prost}. \texttt{Pr\"ost} calculates the posterior probability that each galaxy in a given search region is the true host galaxy using the fractional offset, redshift, and brightness of the host/transient. 
We use a search radius of 60'' and consider galaxies in the \textit{Galaxy List for Advanced Detector Era} catalog (GLADE; \citealt{GLADE}), Panoramic Survey Telescope and Rapid Response System (Pan-STARRS; \citealt{PanSTARRS}) Data Release 2 (DR2; \citealt{Flewelling_2020}), and the DeCaLS Data Release 10. We also flag nuclear transients using \texttt{iinuclear\footnote{\url{https://github.com/gmzsebastian/iinuclear}}}
\citep{iinuclear}, which determines whether the location of the transient coincides with the center of its host galaxy with sufficient probability. In our final candidate selection, we perform human vetting to select the most promising candidates and retain only transients, both nuclear and non-nuclear, with confident host associations.

Our criteria yield 39 transient candidates: 22 in ECDFS and 17 in EDFS. 
\corr{The detections for these candidates are separated by at least 20 days, which effectively excludes moving objects.}
These candidates are listed in Table~\ref{tab:ecdfs_transients} and Table~\ref{tab:edfs_transients}, respectively.

\begin{figure*}
    \centering
    \includegraphics[width=0.49\linewidth]{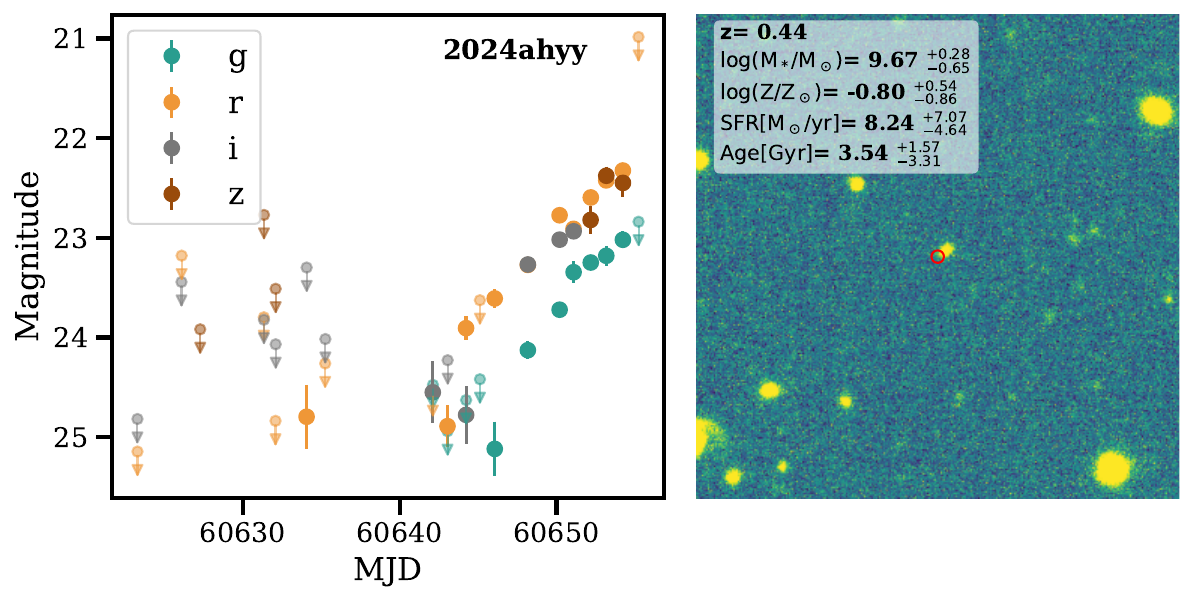}%
    \hfill
    \includegraphics[width=0.49\linewidth]{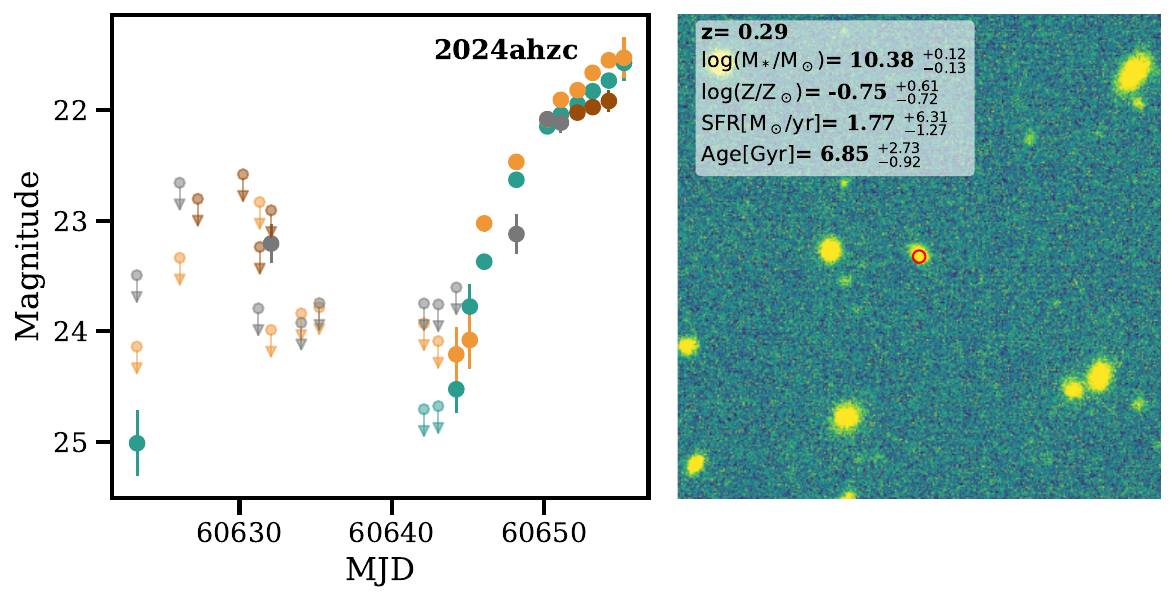}\\[0.5em]
    \includegraphics[width=0.49\linewidth]{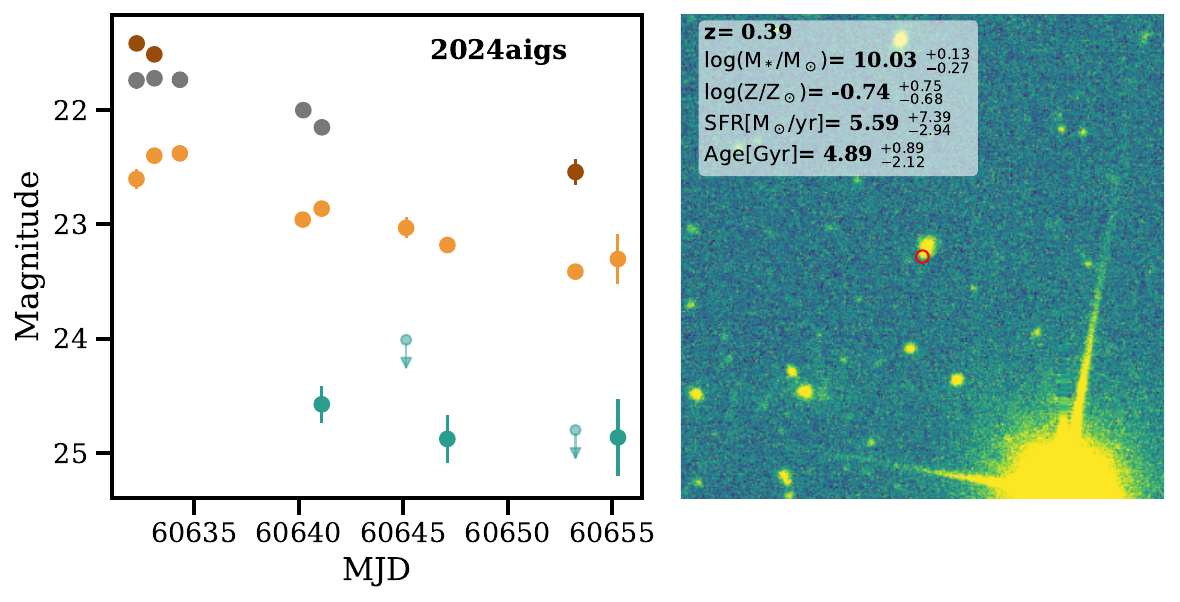}
    \hfill
    \includegraphics[width=0.49\linewidth]{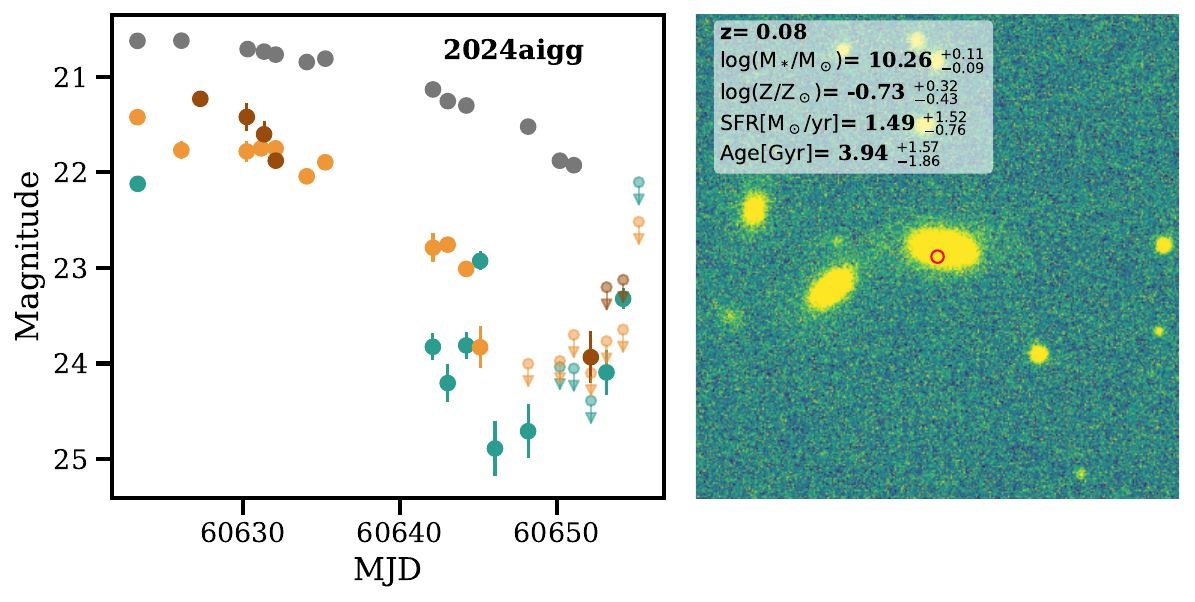}\\[0.5em]
    \includegraphics[width=0.49\linewidth]{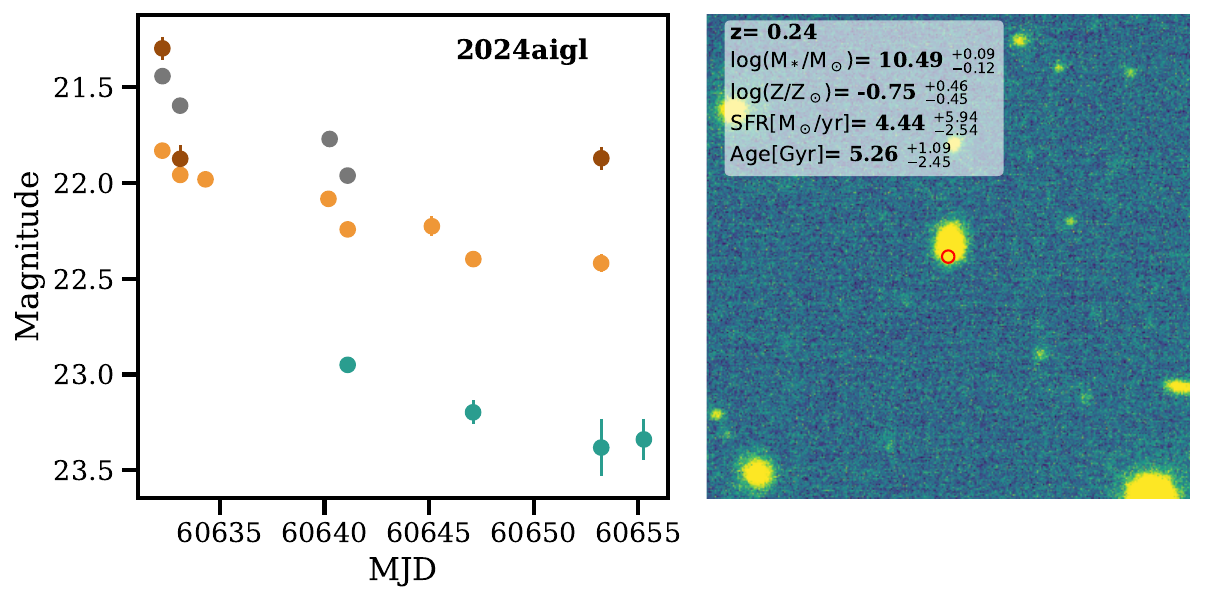}
    \hfill
    \includegraphics[width=0.49\linewidth]{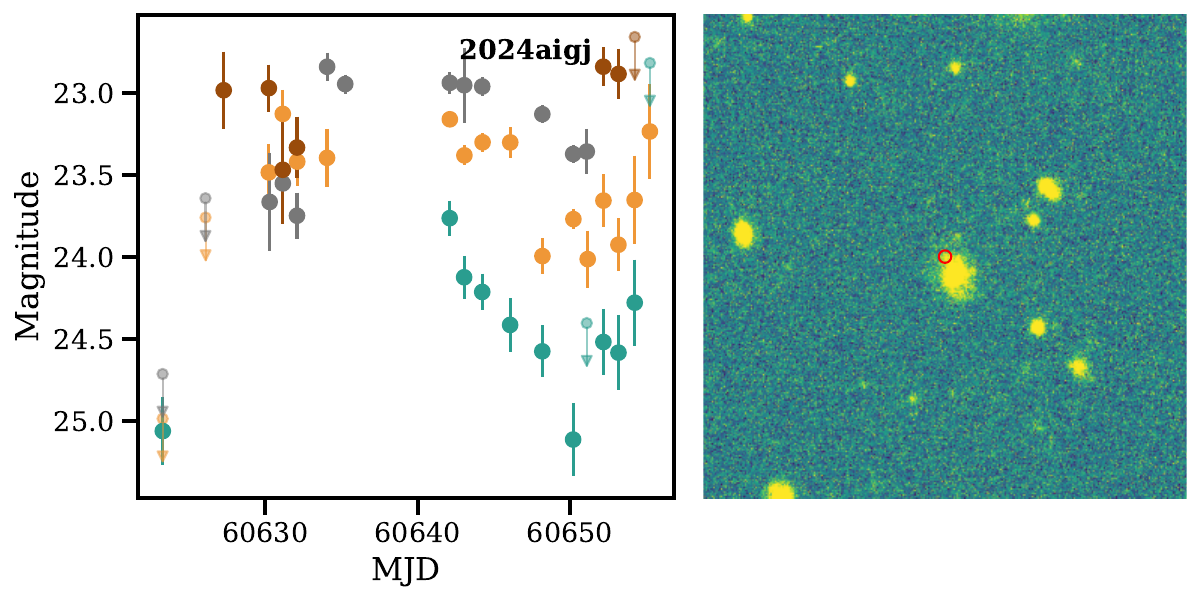}\\[0.5em]
    \includegraphics[width=0.49\linewidth]{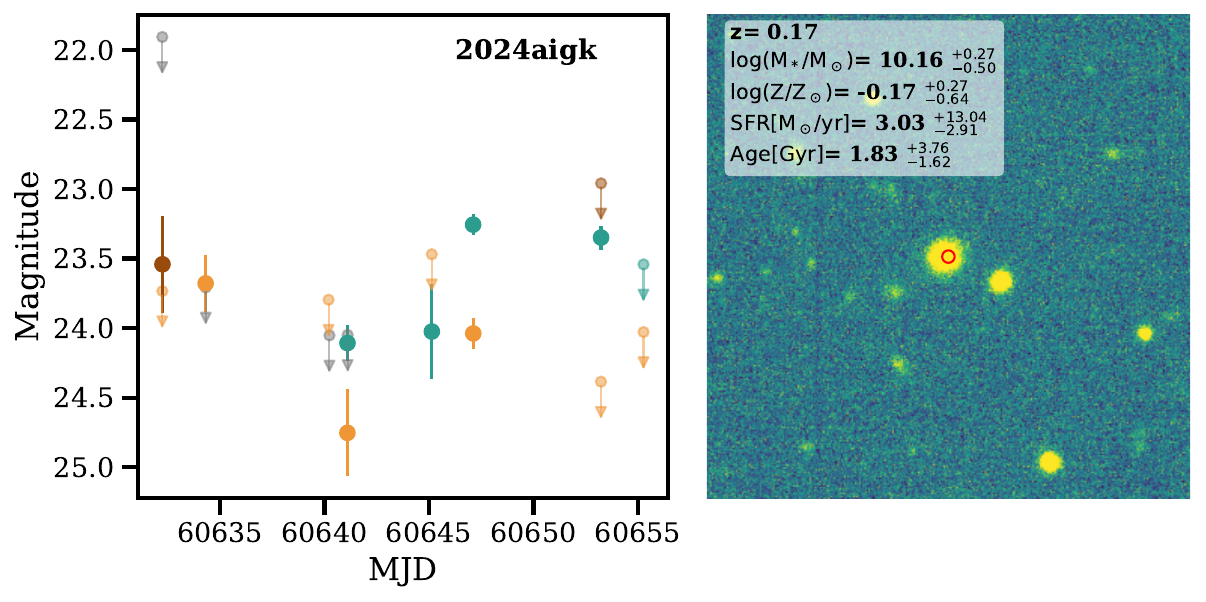}
    \hfill
    \includegraphics[width=0.49\linewidth]{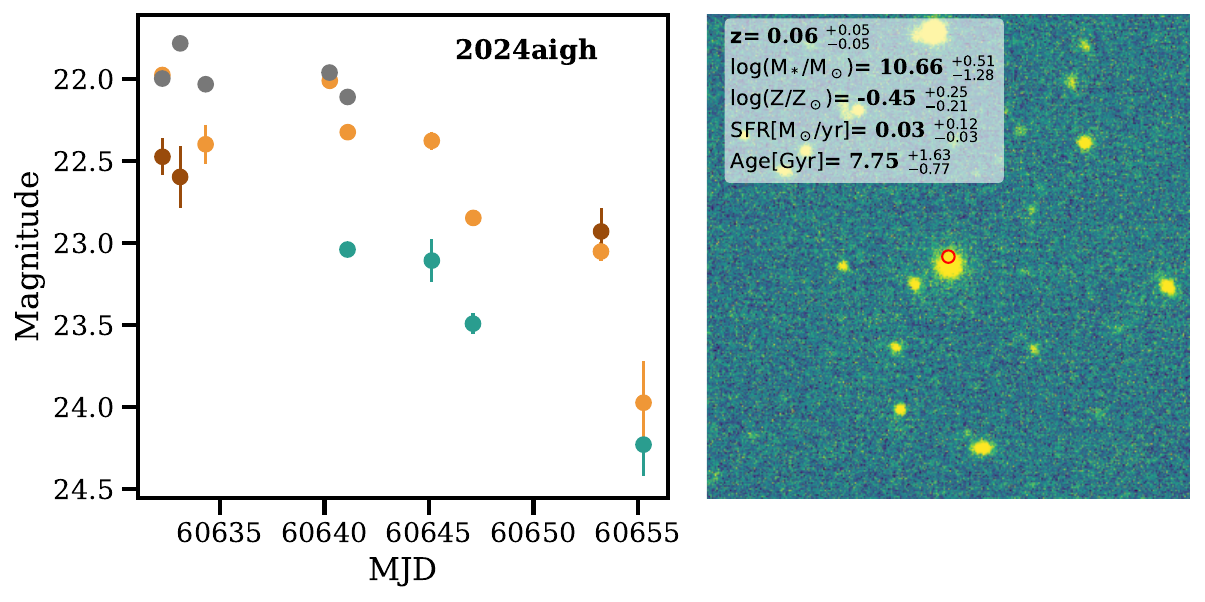}
    \caption{Light curves and host properties of transients previously reported to TNS. Note that plotted errors represent statistical uncertainty. \corr{Host properties include redshift ($z$), stellar mass [$\log(M_{\star}/M_{\odot})$], stellar metallicity [$\log(Z/Z_{\odot})$], star formation rate (SFR; [$M_{\odot},\rm yr^{-1}$]), and mass-weighted stellar population age (Age; [Gyr]).
    }
    }
    \label{fig:candidates}
\end{figure*}

\begin{figure*}
    \centering
    \includegraphics[width=0.49\linewidth]{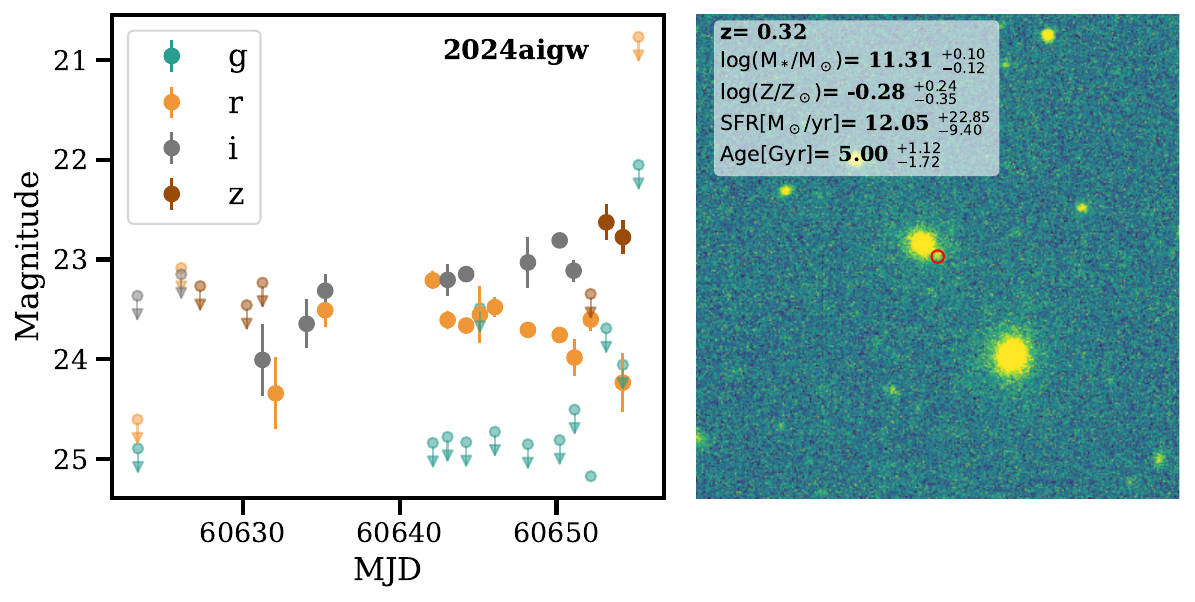}%
    \hfill
    \includegraphics[width=0.49\linewidth]{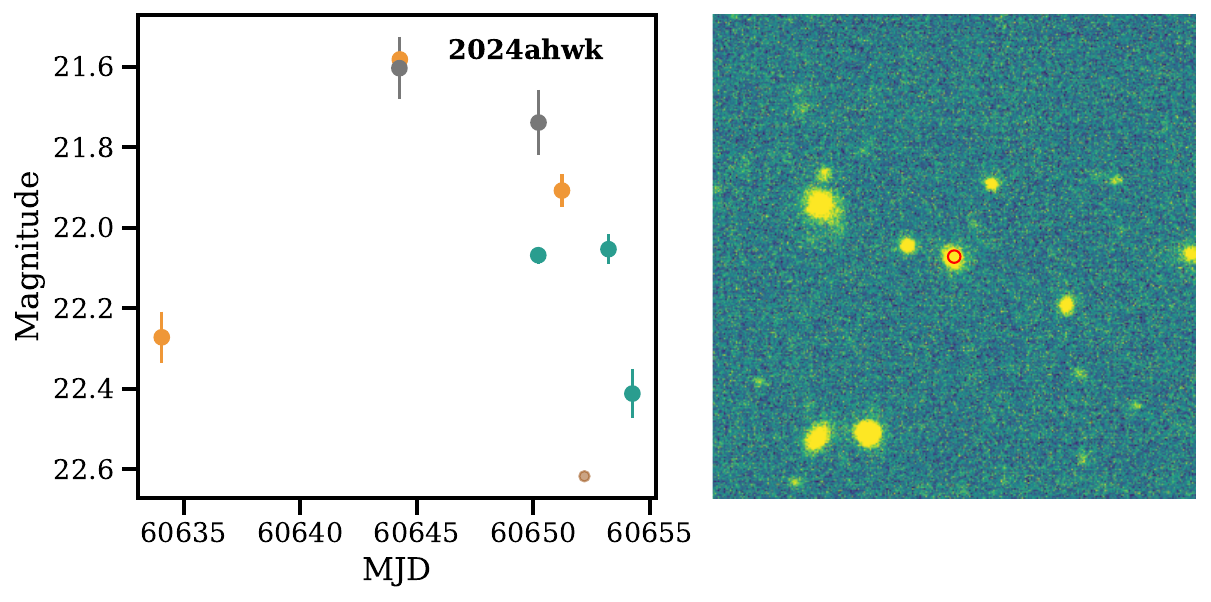}\\[0.5em]
    \includegraphics[width=0.49\linewidth]{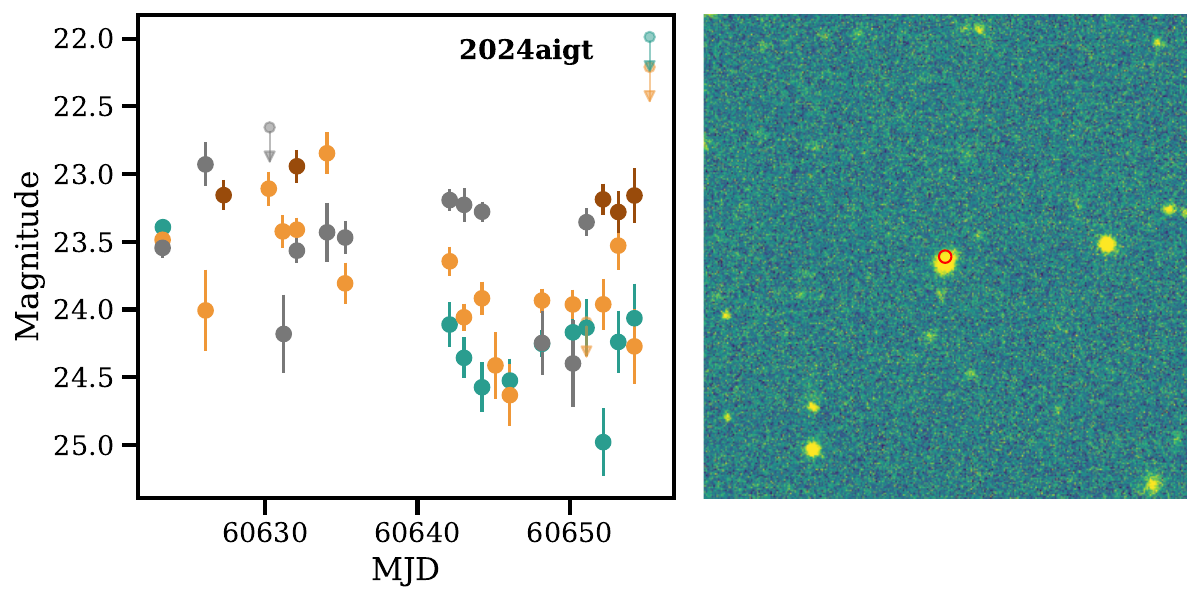}
    \hfill
    \includegraphics[width=0.49\linewidth]{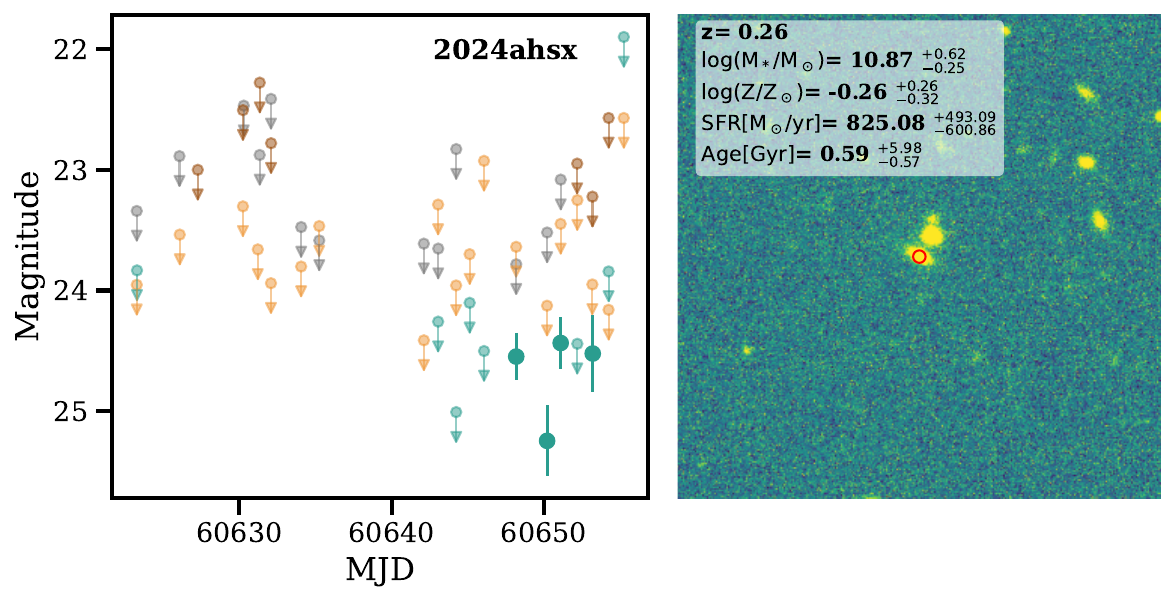}\\[0.5em]
    \includegraphics[width=0.49\linewidth]{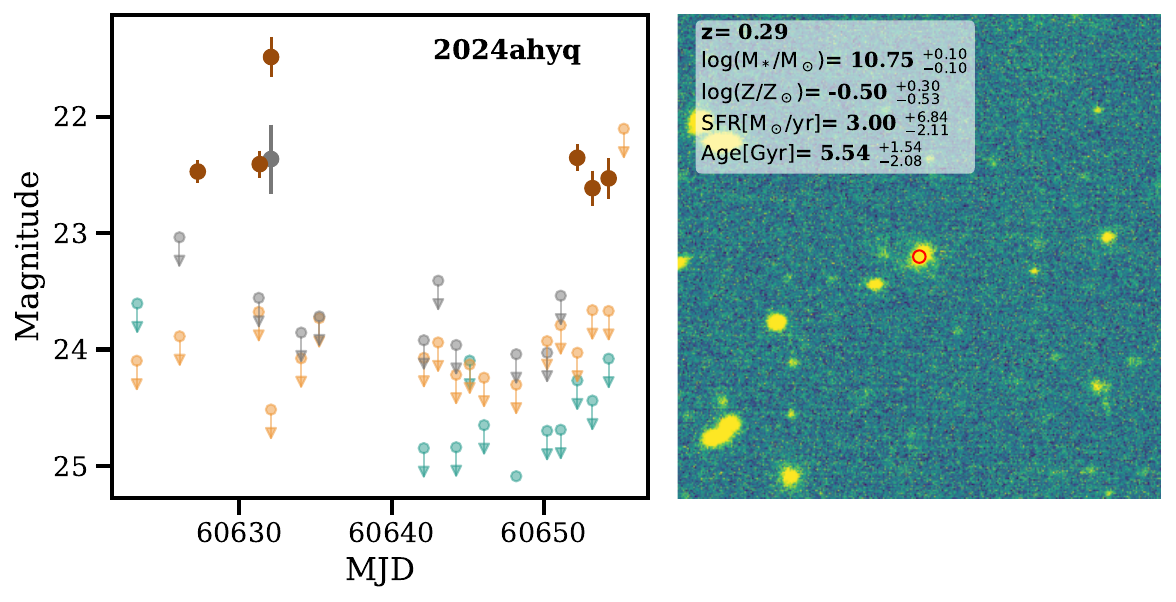}
    \hfill
    \includegraphics[width=0.49\linewidth]{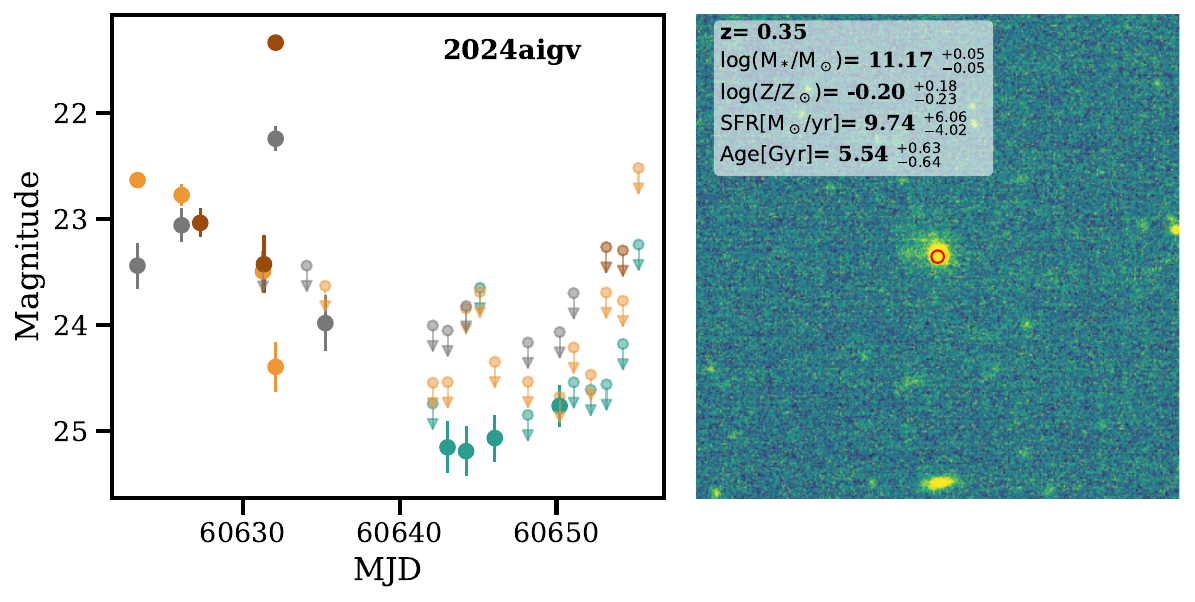}\\[0.5em]
    \caption{Continuation of Figure \ref{fig:candidates} for the remaining transients.}
    \label{fig:candidates_2}
\end{figure*}

\begin{figure*}
    \includegraphics[width=1.\linewidth]{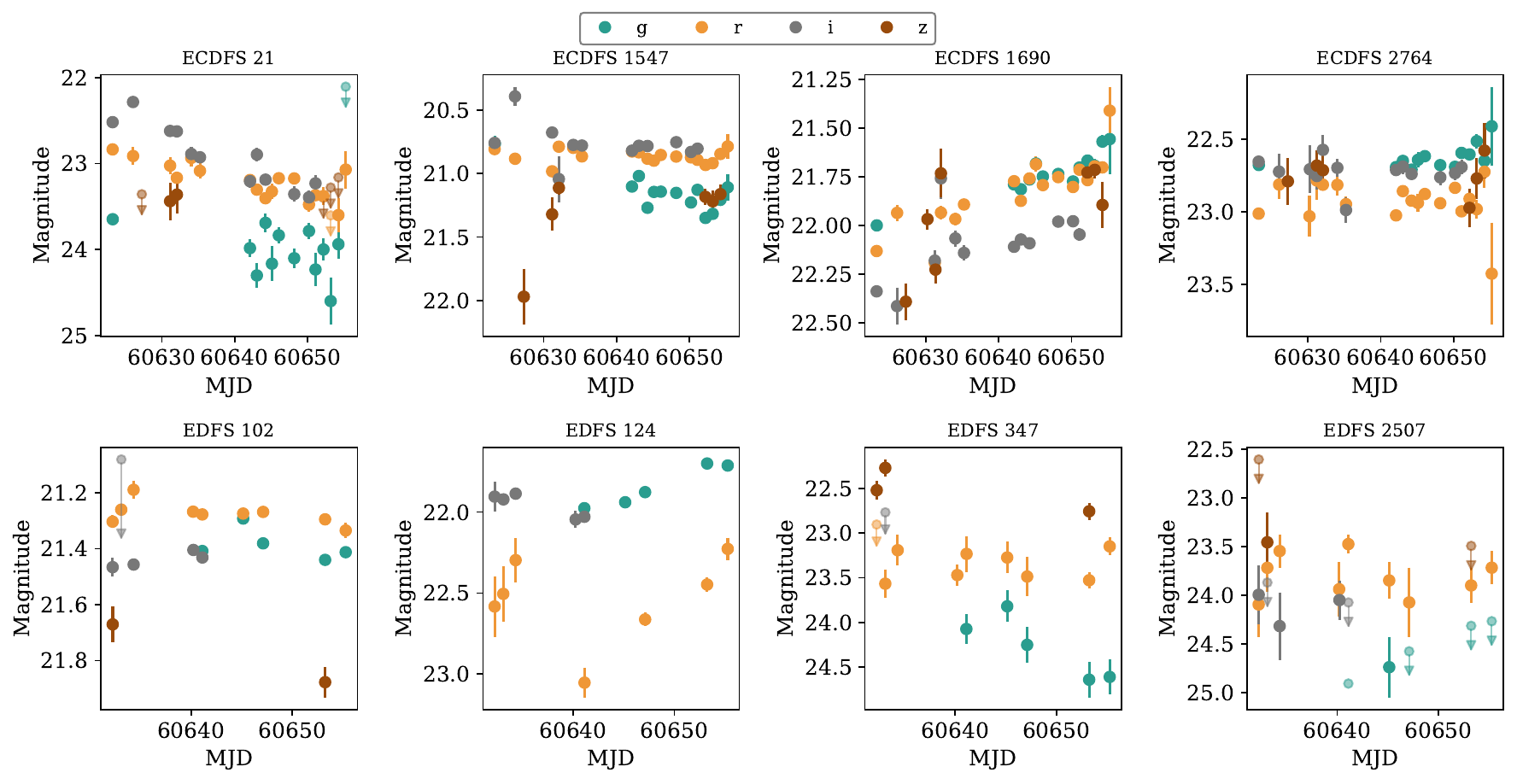}
    \caption{Multiband light curves of a subset of the transient candidates identified by our pipeline. More information about these transients can be found in Table \ref{tab:ecdfs_transients} and \ref{tab:edfs_transients}.
    \label{fig:other_candidates}}
\end{figure*}

\subsection{A Sample of Transient Light Curves in DP1}

\subsubsection{Known Transients}

\cite{Andreoni2025TNSAN.204....1A} reported three newly identified extragalactic transient candidates and eight previously reported transients as visible in the DP1 images. These transients were required to have confident host galaxy associations and to not be co-located with star-like objects or galactic nuclei. Among the 7 reported transients in the ECDFS/EDFS fields, AT~2024aigg \citep{tns_2024aigg}, AT~2024ahzc \citep{tns_2024ahyy_2024ahzc_2024ahzi_2024ahwk_2024ahyq_2024ahsx}, AT~2024ahyy \citep{tns_2024ahyy_2024ahzc_2024ahzi_2024ahwk_2024ahyq_2024ahsx}, and AT~2024aigk \citep{tns_2024aigk_2024aigl} are successfully identified by our search algorithm. AT~2024ahzi \citep{tns_2024ahyy_2024ahzc_2024ahzi_2024ahwk_2024ahyq_2024ahsx}, AT~2024aigl \citep{tns_2024aigk_2024aigl}, and AT~2024ahwk \citep{tns_2024ahyy_2024ahzc_2024ahzi_2024ahwk_2024ahyq_2024ahsx} are not identified because they were not covered by the selected visits used in the transient search. Had they been observed during the visits we selected, our algorithm would have robustly identified them. 

We cross-match the remaining candidates with TNS, and we identify eight additional reported transients: six in the ECDFS field and two in the EDFS field. These transients are AT~2024aigs \citep{tns_2024aigs}, AT~2024aigh \citep{tns_2024aigh}, AT~2024aigt \citep{tns_2024aigt}, AT~2024aigw \citep{tns_2024aigw_2024aigv}, AT~2024aigj \citep{tns_2024aigj}. We also note that AT~2024aigv \citep{tns_2024aigw_2024aigv}, AT~2024ahyq \citep{tns_2024ahyy_2024ahzc_2024ahzi_2024ahwk_2024ahyq_2024ahsx}, and AT~2024ahsx \citep{tns_2024ahyy_2024ahzc_2024ahzi_2024ahwk_2024ahyq_2024ahsx} lie within the field we selected but are not detected, as they has fewer than 3 detections in our selected visits. \corr{We note that the photometric classifications of these TNS transients have been discussed in \citet{Freeburn2025arXiv250722864F}, and refer the reader to that work for further details.}

We perform PSF photometry using \texttt{SLIDE} at the positions of AT~2024ahyy, AT~2024ahzc, AT~2024aigs, AT~2024aigg, AT~2024aigl, AT~2024aigj, AT~2024aigk, AT~2024aigh, AT~2024aigw, AT~2024ahwk, AT~2024aigt, AT~2024ahsx,  AT~2024ahyq, and AT~2024aigv. We refer the reader to \cite{de_soto_in_prep} for photometric analysis of AT~2024ahzi.
\corr{Our transient search uses a subset of $r$-band images, which may not always optimally cover individual objects. To obtain multiband and better temporal coverage for specific objects, we search all DP1 images that overlap with each object's position and select up to two images per night per filter (to reduce computational cost). We then generated difference images and extracted photometry from 1500$\times$1500 pixel cutouts centered on each object using \texttt{SLIDE} across all filters.} 
\corr{Since there are no $u$-band observations in DES and few $y$-band observations in DP1, we performed image subtraction only on the $griz$-band images.}
The light curves and host properties of these objects are shown in Figure~\ref{fig:candidates} and \ref{fig:candidates_2}.

\subsubsection{Unreported Transients}
We cross-match the remaining 29 candidates with the \texttt{DiaObject} table on the Rubin Science Platform, considering only \texttt{DiaObject}s with at least two detections. 17 transients have a corresponding \texttt{DiaObject} within 1 arcsecond. The remaining 12 transients are \corr{either missing or have less than two detections} in the DP1 DIA catalog, because transient flux is contaminating the templates and pushing the difference flux variation below the detection threshold. This highlights the importance of robust templates in transient identification. We show examples of transients with and without \texttt{DiaObject} object associations in Figure \ref{fig:other_candidates}.

\subsection{Host Properties of Transient Candidates} \label{sec:host_property}

In this Letter, we do not perform detailed light curve analysis on our identified candidates; however, as a demonstration of the future workflow in the LSST era, we derive the host properties of each transient using \texttt{FrankenBlast} \citep{FrankenBlast}, a customized version of \texttt{Blast} \citep{Jones2024}. \texttt{FrankenBlast} collects all available images of the host galaxies within the \textit{Galaxy Evolution Explorer} (GALEX; \citealt{Galex}), Pan-STARRS1, DECaLS Data Release 9, \textit{Two Micron All-Sky Survey} (2MASS; \citealt{2MASS}), and \textit{Wide-field Infrared Survey Explorer} (WISE; \citealt{WISE}). It then performs aperture photometry using elliptical apertures constructed for each image, using the \texttt{astropy.photutils} Python package \citep{photutils}. If the host is not detected in a given filter, \texttt{FrankenBlast} adjusts the aperture size using a neighboring filter. Additional technical details are provided in Appendix \ref{appendix:frankenblast}. Host properties of the transient candidates are presented in Table~\ref{tab:ecdfs_transients} and Table~\ref{tab:edfs_transients}, as well as in Figure~\ref{fig:candidates} and Figure~\ref{fig:candidates_2}. Host properties, especially photometric redshifts, can significantly improve the accuracy of photometric classifiers, particularly at early times \corr{\citep{Muthukrishna2019PASP..131k8002M,Boone2019AJ....158..257B,Gomez2020ApJ...904...74G,Sanchez2021AJ....161..141S,gagliano2023first, Kisley2023ApJ...942...29K, superphot_plus, villar2024impact, Sheng2024MNRAS.531.2474S, splash, Gupta2025MNRAS.542L.132G}.}

\section{Discussion and Conclusions} \label{sec:conclusion}
We present the \texttt{SLIDE} package, which performs LSST image subtraction using DECam templates. To demonstrate the anticipated workflow for the LSST survey, we conduct an experimental transient search via the Rubin Science Platform using DP1 difference images produced by this package. We present multiband photometry and host galaxy properties for the most promising transients.

This Letter demonstrates the potential of \texttt{SLIDE} to uncover extended pre-explosion or long-duration transient activity within LSST. In recent years, it has been found that many CCSNe, such as normal SNe~II, SNe~IIn, and SNe~Ibn, interact with dense circumstellar material (CSM) around their progenitors \citep{Khazov2016, Bruch2021, Bruch2022, Leonard2000ApJ...536..239L, Smith2015MNRAS.449.1876S, Yaron2017, Terreran2022ApJ...926...20T, Morozova2020, Wang2020ApJ...900...83W,Gangopadhyay2020ApJ...889..170G,Wang2021ApJ...917...97W,Pellegrino2022ApJ...926..125P, Ben-Ami2023ApJ...946...30B, Bostroem2023ApJ...956L...5B, Pearson2023ApJ...945..107P,Bostroem2023ApJ...953L..18B,Hosseinzadeh2023ApJ...953L..16H,Shrestha2024ApJ...972L..15S,Andrews2024ApJ...965...85A,Meza-Retamal2024ApJ...971..141M,Jacobson-Galan2024ApJ...970..189J,Dong2024ApJ...977..254D,Shrestha2024ApJ...961..247S,Wang2025arXiv250615139W,Gangopadhyay2025arXiv250610700G,Andrews2025ApJ...980...37A, Brennan2025arXiv250308768B}, likely produced months to years before their final explosions. The origin of this CSM, its geometry, and its implications for CCSN progenitor systems remain hotly debated \citep{Chevalier2012, Quataert2012, Soker2013, Shiode2014, Smith2014, Fuller2017, Morozova2020, Metzger22,Wu22,Dessart2022A&A...658A.130D,Tsuna2024arXiv240102389T,Tsuna2024OJAp....7E..82T}.

Improved characterization of precursor emission provides critical insights into the final stages of stellar evolution, and may serve as an early warning signal for imminent CCSNe \citep{Tsuna2023}. To date, precursor activity has been most commonly observed in SNe~IIn \citep{Strotjohann2021ApJ...907...99S,Farias2024ApJ...977..152F,Gagliano2025arXiv250219469G,Pastorello2025arXiv250323123P,Brennan2025arXiv250308768B}. In contrast, only three SNe~Ibn \citep{Pastorello2007Natur.447..829P,Strotjohann2021ApJ...907...99S,Dong2024ApJ...977..254D,Brennan2024A&A...684L..18B} and a single SN~II \citep{Jacobson-Galan2022ApJ...924...15J} have shown evidence for precursor emission, potentially due to their fainter intrinsic luminosities. Moreover, precursor spectroscopy, critical for probing progenitor systems, has only been published for a few events \citep{Pastorello2013ApJ...767....1P,Brennan2024A&A...684L..18B}.
Building a larger sample of SNe with detected precursor emission and precursor spectra is essential for constraining their occurrence rates and understanding their underlying physical mechanisms.

Furthermore, LSST is expected to drastically increase the number of photometrically identified, long-duration precursor events. With a single LSST visit, precursors from normal SNe~II and SNe~IIn can be detected to approximately 100 and 300 Mpc, respectively \citep{Jacobson-Galan2022ApJ...924...15J,Gagliano2025ApJ...978..110G}, while precursors of SNe~Ibn can be detected to approximately 150~Mpc \citep[e.g.,][]{Dong2024ApJ...977..254D}.
Therefore, precursor searches can be constrained to nearby, bright, low-extinction galaxies to reduce computational load and enable rapid identification and follow-up of promising candidates. 
Host association (e.g., with algorithms such as \texttt{Pr\"ost} as we described in Section \ref{sec:selection}) is essential for targeted searches to further decrease the computational load on the RSP. Host galaxy properties can be derived using \texttt{Blast} and \texttt{FrankenBlast}, further enabling prioritization and classification.
We expect \texttt{SLIDE}, paired with the approach outlined in this Letter, to be particularly effective in discovering these long-duration precursors.

Our package will support the study of all transients with slowly evolving phases. LRNe \citep{Mauerhan2015MNRAS.447.1922M,Smith2016MNRAS.458..950S,Blagorodnova2017ApJ...834..107B}, for example, are generally understood to be the product of common envelope episodes and, potentially, mergers \citep[e.g.,][]{Soker03,Tylenda11,Metzger_Pejcha17,Soker2024Galax..12...33S}. LRNe often undergo gradual brightening that lasts for years prior to the main outburst. Multiband photometric and spectroscopic observations during the pre-outburst brightening phase can offer valuable information about their progenitor systems and mass transfer mechanisms preceding the transient \citep{Addison2022MNRAS.517.1884A}. LSST is expected to observe $\sim400-800$ LRNe annually \citep{howitt2020luminous}, drastically increasing the current sample of such events.

Similarly, some extremely energetic transients often evolve slowly.
Superluminous SNe (SLSNe), in particular, are a class of massive stellar explosions with luminosities significantly higher than those of normal SNe, requiring additional power sources beyond radioactive decay \citep{Chomiuk2011ApJ...743..114C,Quimby2011Natur.474..487Q,Gal-Yam2009Natur.462..624G,Howell2013ApJ...779...98H,Gal-Yam2012Sci...337..927G,Howell2017hsn..book..431H,Moriya2018SSRv..214...59M,Gomez2024MNRAS.535..471G}. LSST is expected to discover $\sim10,000$ hydrogen-poor superluminous events annually, with most at high-redshift ($z>1$; \citealt{villar2018superluminous}). Similarly, ambiguous nuclear transients (ANTs) are energetic transients that are found in the nuclei of their host galaxies
\citep{Wiseman2025MNRAS.537.2024W,Pessi2025arXiv250523731P}. These are seemingly distinct from typical AGN and notably more extreme than ``normal" tidal disruption events, although their origin is still an open question. In both cases, the extended duration of these events, especially at high redshift, implies that they are likely to be implanted in the initial LSST templates. 

Early identification of transient events is essential for spectroscopic followup across their evolution. \texttt{SLIDE} enables both early detection and reliable photometry, even when the transients are embedded in their LSST template images. Because the aforementioned transient classes  are rare, the early years of LSST offer an opportunity to build statistically meaningful samples of such events that will guide strategies for follow-up in the future.



\begin{deluxetable*}{cccccccccc}
\tablecaption{ECDFS Transient Candidates\label{tab:ecdfs_transients}}
\tablewidth{0pt}   
\renewcommand{\arraystretch}{1.0}
\setlength{\tabcolsep}{3pt}
\tabletypesize{\scriptsize}   
\tablehead{
\colhead{ID} & 
\colhead{\makecell{RA\\(hh:mm:ss)}} & 
\colhead{\makecell{Declination\\(dd:mm:ss)}} & 
\colhead{\makecell{Nuc.\\Flag}} & 
\colhead{$z_{\rm host}$} & 
\colhead{DIA Object ID} & 
\colhead{$\log(Z/Z_\odot)$} & 
\colhead{$\log(M_*/M_\odot)$} & 
\colhead{\makecell{Age\\(Gyr)}} & 
\colhead{\makecell{SFR\\($M_\odot$/yr)}}
}
\renewcommand{\arraystretch}{1.3}
\setlength{\extrarowheight}{4pt}
\startdata
\multicolumn{10}{c}{Transients Reported to TNS} \\
\hline \hline
2024ahsx & 03:33:28.07 &  -28:12:54.36 & 1 & 0.261(0.011) & 611253629533291776 & $-0.26^{+0.26}_{-0.32}$ & $10.87^{+0.62}_{-0.25}$ & $0.59^{+5.98}_{-0.57}$ & $825.08^{+493.09}_{-600.86}$ \\
2024ahwk & 03:29:50.944 & -28:13:04.73 & 0 & 0.270(0.013) & 611253973130674268 & - & - & - & - \\
2024ahyq & 03:31:37.65 & -28:20:01.31 & 1 & 0.294(0.040) & 609782139377943168 & $-0.50^{+0.30}_{-0.53}$ & $10.75^{+0.10}_{-0.10}$ & $5.54^{+1.54}_{-2.08}$ & $3.00^{+6.84}_{-2.11}$ \\
2024ahyy & 03:31:34.22 & -28:24:45.37 & 0 & 0.438(0.105) & 609781520902651904 & $-0.80^{+0.54}_{-0.86}$ & $9.67^{+0.28}_{-0.65}$ & $3.54^{+1.57}_{-3.31}$ & $8.24^{+7.07}_{-4.64}$ \\
2024ahzc & 03:31:21.18 & -28:16:47.64 & 0 & 0.290(0.042) & 609782208097419264 & $-0.75^{+0.61}_{-0.72}$ & $10.38^{+0.12}_{-0.13}$ & $6.85^{+2.73}_{-0.92}$& $1.77^{+6.31}_{-1.27}$ \\
2024aigg & 03:32:29.94 & -27:44:23.33 & 0 & 0.069(0.015) & 611255759837069440 & $-0.73^{+0.32}_{-0.43}$ & $10.26^{+0.11}_{-0.09}$ & $3.94^{+1.57}_{-1.86}$ & $1.49^{+1.52}_{-0.76}$ \\
2024aigj & 03:32:51.02 & -27:40:52.60 & 0 & 0.251(0.047) & 611256447031836800 & - & - & - & - \\
2024aigt & 03:33:41.37 & -28:13:24.81 & 0 & 0.296(0.053) & 611253629533290624 & - & - & - & - \\
2024aigw & 03:30:55.57 & -27:51:58.87 & 0 & 0.323(0.011) & 611255210081255575 & $-0.28^{+0.24}_{-0.35}$ & $11.31^{+0.10}_{-0.12}$ & $5.00^{+1.12}_{-1.72}$ & $12.05^{+22.85}_{-9.40}$ \\
2024aigv & 03:32:13.81 & -28:28:14.40 & 0 & 0.375(0.055) & 609788942606139423 & $-0.20^{+0.18}_{-0.23}$ & $11.17^{+0.05}_{-0.05}$ & $5.54^{+0.63}_{-0.64}$ & $9.74^{+6.06}_{-4.03}$ \\
\hline
\multicolumn{10}{c}{Unreported Transients} \\
\hline \hline
13 & 03:31:37.69 & -28:04:10.16 & 0 & 0.132(0.020) & -- & -0.91$^{+0.55}_{-0.47}$ & 10.10$^{+0.09}_{-0.11}$ & 5.87$^{+1.11}_{-1.66}$ & 1.51$^{+2.16}_{-0.80}$ \\
14 & 03:31:35.11 & -28:07:14.42 & 0 & 0.127(0.030) & 611254522886494620 & -1.20$^{+0.58}_{-0.45}$ & 9.83$^{+0.15}_{-0.21}$ & 4.79$^{+1.64}_{-4.10}$ & 1.60$^{+3.01}_{-0.98}$ \\
21 & 03:31:41.65 & -28:05:10.74 & 1 & 0.205(0.056) & 611254454167011721 & -0.70$^{+0.75}_{-0.74}$ & 9.63$^{+0.26}_{-0.68}$ & 5.55$^{+1.86}_{-5.29}$ & 3.46$^{+8.91}_{-2.61}$ \\
100 & 03:31:34.14 & -27:49:59.61 & 0 & 0.464(0.125) & -- & -1.06$^{+0.72}_{-0.61}$ & 10.21$^{+0.21}_{-0.77}$ & 3.35$^{+1.45}_{-3.29}$ & 34.04$^{+37.63}_{-20.92}$ \\
706 & 03:33:32.49 & -27:48:32.00 & 1 & 0.333(0.108) & -- & -0.76$^{+0.66}_{-0.88}$ & 9.64$^{+0.36}_{-0.72}$ & 3.99$^{+1.82}_{-3.87}$ & 7.05$^{+9.50}_{-4.86}$ \\
1071 & 03:31:41.92 & -28:04:22.22 & 0 & 0.149(0.033) & -- & -0.69$^{+0.63}_{-0.79}$ & 9.41$^{+0.22}_{-0.65}$ & 4.82$^{+2.42}_{-4.79}$ & 3.54$^{+11.21}_{-2.98}$ \\
1314 & 03:33:07.70 & -27:53:31.81 & 1 & 0.121(0.036) & -- & -0.85$^{+0.71}_{-0.65}$ & 9.21$^{+0.13}_{-0.20}$ & 6.27$^{+1.42}_{-2.76}$ & 0.32$^{+1.26}_{-0.19}$ \\
1350 & 03:32:49.30 & -27:37:57.07 & 1 & 0.131(0.066) & 611256447031837444 & -- & -- & -- & -- \\
1367 & 03:33:21.09 & -27:39:12.07 & 1 & 0.597(0.298) & 611256378312359976 & -- & -- & -- & -- \\
1547 & 03:32:46.03 & -28:22:32.21 & 1 & 0.621(0.216) & 609788873886662802 & -0.47$^{+0.44}_{-0.53}$ & 10.59$^{+0.25}_{-0.63}$ & 3.74$^{+1.08}_{-3.65}$ & 82.07$^{+91.77}_{-45.82}$ \\
1690 & 03:31:20.77 & -27:56:49.26 & 1 & 0.834(0.118) & 611255210081255450 & -0.37$^{+0.35}_{-0.81}$ & 10.39$^{+0.16}_{-0.25}$ & 3.07$^{+0.70}_{-1.48}$ & 14.58$^{+8.55}_{-5.80}$ \\
1897 & 03:33:09.50 & -27:44:06.89 & 1 & 0.117(0.032) & 611255691117594961 & -- & -- & -- & -- \\
1965 & 03:31:24.22 & -27:53:42.27 & 1 & 0.110(0.055) & 611255210081255504 & -- & -- & -- & -- \\
2764 & 03:31:47.87 & -28:17:00.77 & 1 & 1.130(0.231) & -- & -0.16$^{+0.22}_{-0.34}$ & 11.62$^{+0.10}_{-0.10}$ & 3.08$^{+0.48}_{-1.19}$ & 6.19$^{+18.46}_{-4.86}$ \\
2798 & 03:33:46.01 & -28:20:07.43 & 1 & 0.170(0.038) & -- & -0.75$^{+0.53}_{-0.69}$ & 9.41$^{+0.19}_{-0.32}$ & 4.97$^{+1.61}_{-3.83}$ & 1.16$^{+2.44}_{-0.77}$ \\
\enddata
{\bf Note}: Galaxy properties are not derived for hosts with insufficient photometric data.
\end{deluxetable*}

\begin{deluxetable*}{cccccccccc}
\tablecaption{EDFS Transient Candidates\label{tab:edfs_transients}}
\tablewidth{\textwidth}             
\tabletypesize{\scriptsize}   

\tablehead{
\colhead{ID} & 
\colhead{\makecell{RA\\(hh:mm:ss)}} & 
\colhead{\makecell{Declination\\(dd:mm:ss)}} & 
\colhead{\makecell{Nuc.\\Flag}} & 
\colhead{$z_{\rm host}$} & 
\colhead{DIA Object ID} & 
\colhead{$\log(Z/Z_\odot)$} & 
\colhead{$\log(M_*/M_\odot)$} & 
\colhead{\makecell{Age\\(Gyr)}} & 
\colhead{\makecell{SFR\\($M_\odot$/yr)}}
}
\renewcommand{\arraystretch}{1.3}
\setlength{\extrarowheight}{4pt}
\startdata
\multicolumn{10}{c}{Transients Reported to TNS} \\
\hline \hline
2024aigh & 03:57:17.80 & -48:22:08.30 & 0 & 0.06(0.04) & 592915218690999552 & $-0.45^{+0.25}_{-0.21}$ & $10.66^{+0.51}_{-1.28}$ & $7.75^{+1.63}_{-0.77}$ & $0.03^{+0.12}_{-0.03}$ \\
2024aigk & 03:55:31.82 & -48:27:43.71 & 1 & 0.168(0.030) & 592915356129952000 & $-0.17^{+0.27}_{-0.64}$ & $10.16^{+0.27}_{-0.50}$ & $1.83^{+3.76}_{-1.62}$ & $3.03^{+13.04}_{-2.91}$ \\
2024aigl & 03:59:24.16 & -48:46:50.53 & 0 & 0.225(0.021)  & 592913706862510093 & -0.75$^{+0.46}_{-0.45}$ & 10.49$^{+0.09}_{-0.12}$ & 5.26$^{+1.09}_{-2.45}$ & 4.44$^{+5.94}_{-2.54}$ \\
2024aigs & 03:56:53.23 & -49:06:18.06 & 0 & 0.393(0.147) & 591819074317582336 & $-0.74^{+0.75}_{-0.68}$ & $10.03^{+0.13}_{-0.27}$ & $4.89^{+0.89}_{-2.12}$ & $5.59^{+7.39}_{-2.94}$ \\
\hline
\multicolumn{10}{c}{Unreported Transients} \\
\hline \hline
102 & 03:57:09.25 & -48:47:02.21 & 1 & 0.987(0.105) & 592913844301464254 & -0.02$^{+0.15}_{-0.25}$ & 10.96$^{+0.40}_{-0.51}$ & 1.17$^{+1.67}_{-1.09}$ & 416.87$^{+291.91}_{-207.38}$ \\
124 & 03:56:23.21 & -48:21:48.14 & 1 & 0.637(0.076) & 592915287410475027 & -0.25$^{+0.24}_{-0.29}$ & 11.12$^{+0.31}_{-0.54}$ & 3.68$^{+1.32}_{-3.62}$ & 281.54$^{+329.42}_{-183.53}$ \\
199 & 03:56:48.54 & -48:19:13.44 & 1 & 0.148(0.017) & 592915974605242521 & -- & -- & -- & -- \\
206 & 03:57:28.37 & -48:27:49.38 & 0 & 0.156(0.040) & 592915218690998834 & -0.29$^{+0.24}_{-0.35}$ & 10.85$^{+0.06}_{-0.07}$ & 6.41$^{+0.86}_{-0.98}$ & 2.51$^{+2.75}_{-1.31}$ \\
347 & 03:54:35.18 & -48:43:37.65 & 0 & 0.367(0.141) & 592914050459894637 & -0.45$^{+0.40}_{-0.85}$ & 10.53$^{+0.20}_{-0.29}$ & 5.17$^{+1.51}_{-4.16}$ & 7.61$^{+20.32}_{-5.86}$ \\
357 & 03:54:15.75 & -48:35:30.15 & 0 & 1.112(0.202) & 592914737654661956 & 0.07$^{+0.09}_{-0.20}$ & 10.73$^{+0.26}_{-0.14}$ & 0.04$^{+1.68}_{-0.03}$ & 842.01$^{+317.67}_{-244.37}$ \\
392 & 03:57:16.36 & -48:51:04.92 & 1 & 0.829(0.240) & 592913157106705461 & -1.24$^{+0.45}_{-0.34}$ & 11.29$^{+0.14}_{-0.27}$ & 0.51$^{+0.64}_{-0.25}$ & 133.94$^{+204.08}_{-75.63}$ \\
490 & 03:55:53.49 & -48:22:19.23 & 1 & 0.574(0.205) & 592915356129953233 & -- & -- & -- & -- \\
569 & 03:57:27.61 & -48:28:38.66 & 0 & 0.406(0.159) & -- & -- & -- & -- & -- \\
668 & 03:56:41.15 & -48:16:57.78 & 0 & 0.171(0.056) & -- & -0.84$^{+0.63}_{-0.72}$ & 10.34$^{+0.19}_{-0.31}$ & 5.17$^{+1.61}_{-3.18}$ & 2.13$^{+4.01}_{-1.58}$ \\
1236 & 03:58:16.51 & -48:27:33.57 & 1 & 0.399(0.166) & -- & -0.61$^{+0.45}_{-1.07}$ & 9.68$^{+0.20}_{-0.30}$ & 5.16$^{+0.89}_{-1.45}$ & 2.56$^{+1.61}_{-1.05}$ \\
2507 & 03:56:00.29 & -48:45:22.26 & 0 & 0.119(0.059) & -- & -0.67$^{+0.60}_{-0.86}$ & 8.99$^{+0.22}_{-0.31}$ & 5.65$^{+2.05}_{-4.96}$ & 0.14$^{+0.51}_{-0.12}$ \\
4777 & 03:55:53.52 & -49:07:29.71 & 1 & 0.147(0.025) & 591819143037059212 & -- & -- & -- & -- \\
5575 & 03:54:41.14 & -48:34:11.92 & 0 & 0.187(0.037) & -- & -0.68$^{+0.44}_{-0.67}$ & 9.84$^{+0.18}_{-0.31}$ & 5.18$^{+1.57}_{-3.80}$ & 4.96$^{+9.11}_{-3.17}$ \\
\enddata
\end{deluxetable*}

\section*{Acknowledgments}
We thank Griffin Hosseinzadeh for providing the PyZOGY image subtraction example. We thank Gautham Narayan, Tanner Murphey, and Qinan Wang for helpful discussions. We also thank Padma Venkatraman for testing the package and providing valuable feedback.

The Villar Astro Time Lab acknowledges support through the David and Lucile Packard Foundation, the Research Corporation for Scientific Advancement (through a Cottrell Fellowship), the National Science Foundation under AST-2433718, AST-2407922 and AST-2406110, as well as an Aramont Fellowship for Emerging Science Research.  This work is supported by the National Science Foundation under Cooperative Agreement PHY-2019786 (the NSF AI Institute for Artificial Intelligence and Fundamental Interactions).  K.d.S. thanks the LSST-DA Data Science Fellowship Program, which is funded by LSST-DA, the Brinson Foundation, the WoodNext Foundation, and the Research Corporation for Science Advancement Foundation; her participation in the program has benefited this work.

This research made use of Photutils, an Astropy package for detection and photometry of astronomical sources \citep{larry_bradley_2022_6825092}.

This research uses services or data provided by the Astro Data Lab, which is part of the Community Science and Data Center (CSDC) Program of NSF NOIRLab. NOIRLab is operated by the Association of Universities for Research in Astronomy (AURA), Inc. under a cooperative agreement with the U.S. National Science Foundation.

This work has made use of data from the European Space Agency (ESA) mission
{\it Gaia} (\url{https://www.cosmos.esa.int/gaia}), processed by the {\it Gaia}
Data Processing and Analysis Consortium (DPAC,
\url{https://www.cosmos.esa.int/web/gaia/dpac/consortium}). Funding for the DPAC
has been provided by national institutions, in particular the institutions
participating in the {\it Gaia} Multilateral Agreement.

\facilities{ADS, Astro Data Lab, Rubin:Simonyi(LSSTComCam), Rubin:USDAC
}

\software{Astropy \citep{astropy13,astropy18, Astropy2022ApJ...935..167A}, 
\texttt{Pr\"ost} \citep{Gagliano2025_Prost},
          Matplotlib \citep{Hunter2007},
          NumPy \citep{2020Natur.585..357H},
          Pandas \citep{mckinney-proc-scipy-2010},
          SciPy \citep{2020NatMe..17..261V},
          Butler \citep{Jenness2022SPIE12189E..11J},
          Photutils \citep{photutils}
          }

\appendix
\section{\texttt{FrankenBlast} Stellar Population modeling} \label{appendix:frankenblast}
\texttt{FrankenBlast} constrains host galaxy stellar population properties using \texttt{SBI++}, a simulation based inference technique that learns posterior density distributions of stellar population properties from simulated galaxy photometry \citep{wlv+2023}. \texttt{FrankenBlast} trains its model on 2 million simulated galaxies from \texttt{Prospector} \citep{Leja2019, jlc+2021}, a stellar population modeling inference code, which uses \texttt{FSPS} and \texttt{python-FSPS} \citep{FSPS_2009, FSPS_2010} to create mock photometry from a given set of stellar population properties. The mock photometry is noised up to match the SNR of the observed sources within the aforementioned surveys used for photometry. \texttt{FrankenBlast} employs the \citet{Chabrier2003} initial mass function (IMF), the \citet{KriekandConroy13} dust attenuation model and \citet{DraineandLi07} IR dust extinction model, a nebular emission model \citep{bdc+2017}, and an AGN mid-IR model within its \texttt{Prospector} model. It tracks the star formation history (SFH) of the hosts through a seven-bin non-parametric model. This SFH model assumes a constant star-formation rate (SFR) in a single age-bin: the first two age-bins are linearly spaced from 0-30~Myr and 30-100~Myr, and the final five are log-spaced until the age of Universe at the host's redshift (or sampled redshift, if it is not known). We report present-day SFR as the SFR in the first two age bins. For accurate constraints on stellar metallicity, \texttt{FrankenBlast} includes the \citet{gcb+05} mass-metallicity relation. When fitting the observed host galaxy data, we have the option to set the host's redshift at a specific value (spec-$z$ model), or sample redshift as a fit parameter (photo-$z$ model; assumes $0 \leq z \leq 1.5$). If a spectroscopic or photometric redshift estimate of the host galaxy is given from the \texttt{Pr\"ost} host association (see Section \ref{sec:fields}), we use that redshift as the redshift of the host and fit with the spec-$z$ model. Otherwise, redshift is determined through the photo-$z$ model.


\bibliography{lsst_DECam}{}
\bibliographystyle{aasjournal}



\end{CJK*}
\end{document}